\PassOptionsToPackage{hyphens}{url}
\documentclass[camera,letterpaper,nomarginnotes,nonarrowgutter]{jpaper}

\usepackage[sort,compress]{cite}
\usepackage{amsmath,amssymb}
\usepackage{graphicx}
\usepackage{xspace}
\usepackage{booktabs}
\usepackage{multirow}
\usepackage{array}
\usepackage{xcolor}
\usepackage{tikz}
\usetikzlibrary{arrows.meta, shadows}
\usepackage[most]{tcolorbox}
\usepackage{pifont}
\newcommand{\cmark}{\ding{51}}
\newcommand{\xmark}{\ding{55}}
\newcommand{\pmark}{\raisebox{-0.1ex}{\scalebox{1.45}{\ensuremath{\triangle}}}}
\newcommand{\tabhead}[1]{\begin{tabular}[b]{@{}c@{}}#1\end{tabular}}

\newcommand{\model}{our model\xspace}
\newcommand{\Model}{Our model\xspace}
\newcommand{\hcfirst}{$\mathrm{HC}_{\mathrm{first}}$\xspace}
\newcommand{\tAggON}{$t_{\mathrm{AggON}}$\xspace}
\newcommand{\tAggOFF}{$t_{\mathrm{AggOFF}}$\xspace}
\newcommand{\NWL}{NWL\xspace}
\newcommand{\PWL}{PWL\xspace}
\newcommand{\vzero}{V0/A1\xspace}
\newcommand{\vone}{V1/A0\xspace}

\newcommand{\head}[1]{\noindent\textbf{\boldmath #1.}~}

\tcbset{before skip=4.1pt, after skip=4.1pt}
\newtcolorbox{tkx}[2][]{%
  enhanced, breakable,
  colframe  = #2!60!black,
  colback   = #2!5,
  boxrule   = 0.6pt,
  arc       = 1.5pt,
  left      = 4pt,
  right     = 4pt,
  top       = 3pt,
  bottom    = 3pt,
  #1%
}
\definecolor{obscolor}{HTML}{E07B39}
\definecolor{takecolor}{HTML}{7B68EE}
\newcounter{obsv}\newcounter{takeaway}
\newcommand{\observation}[1]{\refstepcounter{obsv}%
  \begin{tkx}{obscolor}\noindent\textbf{\sffamily Observation~\theobsv.} #1\end{tkx}}
\newcommand{\takeaway}[1]{\refstepcounter{takeaway}%
  \begin{tkx}{takecolor}\noindent\textbf{\sffamily Takeaway~\thetakeaway.} #1\end{tkx}}
\newcommand*\circled[1]{\tikz[baseline=(c.base)]\node[shape=circle,
  fill=black,text=white,inner sep=0.6pt](c){\scriptsize\textbf{#1}};}

\newcommand{\modname}[3]{Mfr.~#1 #2Gb #3-Die}
\makeatletter
\newcommand{\defmodule}[4]{\expandafter\def\csname modref@#1\endcsname{\modname{#2}{#3}{#4}}}
\newcommand{\modref}[1]{\ifcsname modref@#1\endcsname #1, \csname modref@#1\endcsname%
  \else\PackageError{preamble}{Unknown module ID #1}{Define it with \string\defmodule.}\fi}
\makeatother
\defmodule{S00}{S}{4}{E} \defmodule{S01}{S}{4}{F} \defmodule{S02}{S}{8}{B} \defmodule{S03}{S}{8}{D}
\defmodule{S04}{S}{16}{A} \defmodule{S05}{S}{16}{B} \defmodule{S06}{S}{16}{M}
\defmodule{H00}{H}{4}{A} \defmodule{H01}{H}{8}{A} \defmodule{H02}{H}{8}{C} \defmodule{H03}{H}{8}{D}
\defmodule{H04}{H}{8}{D} \defmodule{H05}{H}{8}{D} \defmodule{H06}{H}{8}{M} \defmodule{H07}{H}{16}{A}
\defmodule{H08}{H}{16}{C}
\defmodule{M00}{M}{4}{B} \defmodule{M01}{M}{16}{B} \defmodule{M02}{M}{16}{E} \defmodule{M03}{M}{16}{F}
\defmodule{N00}{N}{8}{C} \defmodule{N01}{N}{8}{C}
\defmodule{C00}{C}{8}{A} \defmodule{C01}{C}{16}{A}

\newcommand{\RHFundamentalCitations}{\cite{%
  kim2014flipping, kim2020revisiting, orosa2021deeper, cojocar2020rowhammer, mutlu2017rowhammer, mutlu2019rowhammer,
  mutlu2023fundamentally, mutlu2023retrospective}}

\newcommand{\rowHammerGetsWorseCitations}{\cite{%
  kim2020revisiting, frigo2020trrespass, kim2014flipping, mutlu2017rowhammer, mutlu2018rowhammer, mutlu2019rowhammer,
  mutlu2023fundamentally, orosa2021deeper, yaglikci2022understanding}}

\newcommand{\exploitingRowHammerAllCitations}{\cite{%
  frigo2020trrespass, hassan2021utrr, mutlu2019rowhammer, kamadan2025ecc, nazaraliyev2025not, orosa2024spyhammer,
  woo2025mitigations, aga2017good, agarwal2018rowhammer, aydin2022cyber, barenghi2018software, bhattacharya2016curious,
  bhattacharya2018advanced, bolcskei2025rubicon, bosman2016dedup, bostanci2025understanding, brasser2017can, burleson2016invited,
  cai2022feasibility, carre2018openssl, chen2025rhammer, cohen2022hammerscope, cojocar2019eccploit, cojocar2020rowhammer,
  deridder2021smash, deridder2025posthammer, fahr2022effects, fahr2022frodo, fournaris2017exploiting, frigo2018grand,
  genssler2022reliability, google-project-zero, gruss2016rowhammer, gruss2018another, hong2019terminal, hu2026gddrhammer,
  islam2022signature, jang2017sgx, jattke2022blacksmith, jattke2024zenhammer, jattke2025mcsee, ji2019pinpoint,
  kang2024sledgehammer, kaur2022work, khan2018analysis, kim2014flipping, kim2020revisiting, kogler2022half,
  kwong2020rambleed, lefforge2023reverse, li2014write, li2022cyberradar, li2025oneflip, lim2017active,
  lin2025gpuhammer, lin2026gpubreach, lipp2018nethammer, liu2022generating, luo2023rowpress, meyer2026phoenix,
  mus2022jolt, mutlu2017rowhammer, mutlu2023fundamentally, ni2018write, olgun2024read, orosa2021deeper,
  park2016experiments, park2016statistical, pessl2016drama, poddebniak2018attacking, qiao2016new, rakin2022deepsteal,
  razavi2016flip, roohi2022efficient, rowhammergithub, ryu2017overcoming, seaborn2015exploiting, shukla2026prowhammer,
  staudigl2022neurohammer, tatar2018defeating, tatar2018throwhammer, tobah2022spechammer, tol2022toward, van2016drammer,
  van2018guardion, walker2021ondramrowhammer, wan2026geforge, wang2022research, weissman2020jackhammer, xiao2016one,
  yaglikci2022understanding, yang2019trap, yang2022socially, yao2020deephammer, yuksel2025pudhammer, yun2018study,
  zhang2018triggering, zhang2020pthammer, zhang2022implicit, zheng2022trojvit}}

\newcommand{\mitigatingRowHammerAllCitations}{\cite{%
  yaglikci2021blockhammer, park2020graphene, qureshi2022hydra, olgun2024abacus, canpolat2025chronus, fiedler2026memory,
  hajiabadi2025charm, jedec2024ddr5, kim2023how, kim2025per, orosa2024spyhammer, saxena2024rubix,
  saxena2024start, stefanDRFM, taneja2026mirza, wi2026rowarmor, woo2023rampart, woo2025mitigations,
  wu2026apt, AppleRefInc, Nale2019, Qureshi2024MINT, ajorpaz2022evax, arikan2022processor,
  aweke2016anvil, bains-merged, bains14c, bains14d, bains2016row, bennett2021panopticon,
  bock2019riprh, bostanci2024comet, bostanci2025understanding, brasser2017can, canpolat2024breakhammer, canpolat2024understanding,
  cohen2022hammerscope, cojocar2020rowhammer, deridder2021smash, devaux2021method, di2023copy, enomoto2022efficient,
  fahr2022frodo, fakhrzadehgan2022safeguard, france2022modeling, france2022reducing, frigo2020trrespass, greenfield2012throttling,
  guha2022criticality, han2021surround, hassan2021utrr, hassan2024self,
  hong2023dsac, hp2015rowhammer, irazoqui2016mascat, jaleel2024pride, jattke2022blacksmith, joardar2022learning,
  joardar2022machine, juffinger2023csi, kakolyris2026columnkeeper, kang2020cattwo, kim2014flipping, kim2015architectural,
  kim2021hammerfilter, kim2022mithril, kim2023ddr5, kogler2022half, konoth2018zebram, kwong2020rambleed,
  lee2019twice, lee2021cryoguard, lenovo2015rowhammer, lin2015handling, lin2025cnc, liu2022generating,
  loughlin2021stop, loughlin2022moesiprime, lu2025counterpoint, manzhosov2022revisiting, marazzi2022protrr, marazzi2023rega,
  mutlu2023fundamentally, naseredini2022alarm, park2022rowhammer, qazi2025drfm, qureshi2024impress, qureshi2025autorfm,
  qureshi2025moat, qureshi2026salt, rakin2022deepsteal, saileshwar2022randomized, saroiu2022configure, saroiu2022price,
  saroiu2024ddr5, saxena2022aqua, saxena2023pt, seyedzadeh2017counter, seyedzadeh2018cbt, son2017making, taneja2025dream, tobah2022spechammer, tol2022toward, tomita2022extracting, tugrul2025understanding,
  van2018guardion, vig2018rapid, vittal2025mopac, wang2021discreet, weissman2020jackhammer, wi2023shadow,
  woo2023scalable, woo2025dapper, woo2025qprac, yaglikci2021security, yaglikci2022hira, yaglikci2024spatial,
  yao2020deephammer, you2019mrloc, zhang2020leveraging, zhang2020pthammer, zhang2022implicit, zhang2022softtrr,
  zheng2022trojvit, zhou2022ltpim, zhou2023dnndefender}}

\newcommand{\refreshBasedRowHammerDefenseCitations}{\cite{%
  kim2014flipping, park2020graphene, qureshi2022hydra, kim2022mithril, marazzi2022protrr, marazzi2023rega,
  olgun2024abacus, Qureshi2024MINT, qureshi2025moat, canpolat2025chronus, kim2025per, saxena2024start,
  taneja2026mirza, wu2026apt, aweke2016anvil, bains14d, bains-merged, bains2016row,
  bostanci2024comet, canpolat2024breakhammer, canpolat2024understanding, hassan2024self, jedec2024ddr5, kang2020cattwo,
  kim2015architectural, lee2019twice, qureshi2026salt, seyedzadeh2017counter, seyedzadeh2018cbt, vittal2025mopac,
  yaglikci2021security, yaglikci2022hira, yaglikci2024spatial}}

\usepackage{balance}
\AddToHook{env/thebibliography/end}{\balance}
\titleformat{\section}[hang]{\normalfont\large\bfseries\raggedright}{\thesection. }{0pt}{}
\titleformat{\subsection}[hang]{\fontsize{11pt}{\baselineskip}\bfseries\raggedright}{\thesubsection. }{0pt}{}

\usepackage{fancyhdr}
\usepackage[bookmarks=true,breaklinks=true,colorlinks,citecolor=blue,linkcolor=blue,urlcolor=blue]{hyperref}
\hypersetup{pdftitle={A Principled Architecture-Level Error Model for Understanding, Evaluating, and Mitigating DRAM
  Read Disturbance},
  pdfauthor={Haocong Luo, Zhaoqiang Bai, Ataberk Olgun, {\.I}smail Emir Y{\"u}ksel, Nisa Bostanci, Jikun Wang,
    Maria Makeenkova, Onur Mutlu}}

\title{A Principled Architecture-Level Error Model for\\Understanding, Evaluating, and Mitigating\\DRAM Read Disturbance}

\author{
Haocong Luo$^{1}$\quad
Zhaoqiang Bai$^{2}$\quad
Ataberk Olgun$^{1}$\quad
{\.I}smail~Emir~Y{\"u}ksel$^{1}$\\[0.3em]
Nisa Bostanci$^{1}$\quad
Jikun Wang$^{1}$\quad
Maria Makeenkova$^{1}$\quad
Onur Mutlu$^{1}$\\[0.4em]
{\textit{$^{1}$ETH Zurich\quad
$^{2}$Beijing Superstring Academy of Memory Technology}}
}

\begin{document}

\maketitle
\thispagestyle{fancy}

\begin{abstract}
DRAM read disturbance (e.g., RowHammer and RowPress) is a fundamental DRAM security, reliability, and robustness issue because it can cause bitflips in \emph{unaccessed} DRAM rows. Existing architecture-level DRAM read disturbance error models do \emph{not} fully capture and oversimplify the sophisticated conditions (e.g., access patterns, timings, and data patterns) of DRAM read disturbance bitflips. Existing device-level research works do not directly provide a practical methodology to enable architecture- and system-level error modeling of DRAM read disturbance.

Our goal in this paper is to build an \emph{architecture-level} error model for DRAM read disturbance that 1) is principled on device-level error mechanisms, and 2) faithfully reproduces the quantitative vulnerability metric (i.e., the minimum aggressor row activation count to induce at least one bitflip, \hcfirst) of DRAM read disturbance at different access patterns and timings. The key idea of our practical architecture-level modeling is to 1) represent a victim row with three distinct types of victim cell susceptible to different device-level error mechanisms, and 2) model how the three types of victim cells respond to the opening and closing of the two wordlines adjacent to the victim cell (i.e., the neighboring wordline, NWL, and the passing wordline, PWL) through the mechanisms identified by device-level works.

We validate our model on 24 commercial off-the-shelf DDR4 modules with DRAM chips from five different manufacturers and one HBM2 stack. For each module, we numerically fit our model to 72 \hcfirst measurements, and validate against 178 new measurements on 1) new victim rows, and 2) new access patterns and timings, including timings beyond the fitted range. Our results show that 1) 94.4\% of our model's predicted \hcfirst values agree with the validation measurements ($\pm$15\%), with a median error of 5.4\%, 2) our model is as accurate on new victim rows (95.0\%) as at new access patterns and timings (94.7\%), and 3) our model is equally accurate on the HBM2 stack (93.8\%).

We present two use cases that show how architects can use our model to better evaluate and mitigate a system's vulnerability to DRAM read disturbance. First, by integrating our model with gem5 and Ramulator 2, we demonstrate that our model can accurately reproduce the key RowPress bitflip characteristics from a real program used to demonstrate RowPress bitflips in full-system simulation. Second, we show how our model can guide the design and configuration of DRAM read disturbance mitigations to more accurately account for RowPress compared to prior works. We will open-source our model.

\end{abstract}

\section{Introduction}
\label{sec:introduction}

Accessing a DRAM row (i.e., the aggressor row) can cause \emph{bitflips} in other, unaccessed DRAM rows (i.e., victim rows) due to DRAM read disturbance. Two prominent examples of DRAM read disturbance are 1) RowHammer, where the aggressor row(s) are repeatedly opened and closed \emph{many times}~\RHFundamentalCitations{}, and 2) RowPress, where the aggressor row(s) are kept open for a long time~\cite{luo2023rowpress, luo2024rowpress, luo2024experimental}. Because these bitflips break the fundamental security property of memory isolation, DRAM read disturbance is a critical security, reliability, and robustness issue. Many works demonstrate attacks that exploit it~\exploitingRowHammerAllCitations{}, chips become more vulnerable as DRAM technology scales~\rowHammerGetsWorseCitations{}, and thus systems need mitigations against DRAM read disturbance bitflips~\mitigatingRowHammerAllCitations{}.

DRAM read disturbance is a sophisticated phenomenon. Prior works show that DRAM read disturbance bitflips are highly sensitive to the DRAM access pattern, timings, and data patterns in the aggressor and victim rows~\cite{kim2014flipping, kim2020revisiting, orosa2021deeper, luo2023rowpress, luo2024experimental, yaglikci2024spatial, olgun2025variable}. For example, the key DRAM read disturbance vulnerability metric, the minimum aggressor row activation count to induce at least one bitflip, \hcfirst, can differ by 1) on average $4.8\times$ between single- and double-sided RowHammer~\cite{luo2025revisiting, vts25github}, and 2) orders of magnitude between RowHammer and RowPress~\cite{luo2023rowpress}. Prior works~\cite{luo2023rowpress, luo2025revisiting, walker2021ondramrowhammer, zhou2023double, Zhou2024Understanding} show that the RowHammer and RowPress bitflip direction depends on the device-level structure of DRAM cells.

Existing architecture-level error models of DRAM read disturbance oversimplify the conditions of read-disturbance bitflips by \emph{not} taking the access patterns, timings, and data patterns all into account~\cite{france2022modeling, thomas2023hammulator, goswami2026hammersimsystemleveltoolmodel}. Existing device-level works explain the physical mechanisms of RowHammer and RowPress with TCAD simulations~\cite{yang2019trap, walker2021ondramrowhammer, zhou2023double, Zhou2024Understanding, luo2026demystifying, bai2026unified}, but they do not directly provide a practical error modeling methodology at the architecture- and system-level.

\textbf{Our goal} is to build an \emph{architecture-level} error model for DRAM read disturbance that 1) is principled on device-level failure mechanisms, and 2) faithfully reproduces \hcfirst at different DRAM access patterns, timings, and aggressor-victim data patterns. Based on device-level failure mechanisms, our key modeling methodology consists of two parts. First, we abstract a victim row into distinct cell types that react \emph{differently} to the device-level error mechanisms. We distinguish between \emph{tail} cells, which are the most sensitive to RowHammer, and \emph{bulk} cells, which are the most sensitive to RowPress. Further, based on device-level mechanisms that explain the bitflip direction of RowHammer, we distinguish between tail cells that have the aggressor wordline as their \emph{neighboring wordline} (NWL), i.e., \emph{tail \NWL} cells, and those that have the aggressor wordline as their \emph{passing wordline} (PWL), i.e., \emph{tail \PWL} cells. Second, we explicitly model the read disturbance to the victim cell (i.e., voltage change) caused by the opening and closing of the NWL and the PWL, respectively, using closed-form analytical formulation of the underlying device-level physical mechanisms, instead of a simple aggregate hammer count.

\begin{figure}[t]
\centering
\includegraphics{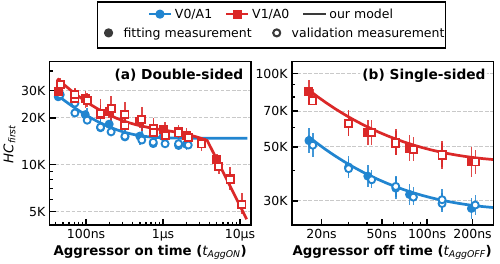}
\caption{\hcfirst predicted by \model (solid lines) vs.\ real-chip measurements that we use to fit (filled markers) and validate (hollow markers) the model, for an example Samsung DDR4 module. (a)~double-sided access pattern at different aggressor row on times (\tAggON) and (b)~single-sided access pattern at different aggressor row off times (\tAggOFF).}
\label{fig:intro-curve}
\end{figure}

Figure~\ref{fig:intro-curve} illustrates an example of how our model predicts \hcfirst (solid lines) for (a)~the double-sided access pattern (left plot) as the aggressor row on time (\tAggON, x-axis) changes, and (b)~the single-sided access pattern (right plot) as the aggressor row off time (\tAggOFF, x-axis) changes, compared to real-chip measurements that we use to numerically fit (filled markers) and validate (hollow markers) the model from an example Samsung module. The blue lines/markers represent the data pattern where the victim cell stores a 0 and the aggressor cell stores a 1 (i.e., the V0/A1 data pattern; red lines/markers vice versa). We observe that our model accurately predicts both the numerical \hcfirst value and how \hcfirst changes as the access pattern, timings, and data pattern change. On this module, 97.2\% of all 178 validation measurements (including other access patterns, timings, and data patterns not shown in~Figure~\ref{fig:intro-curve} for brevity) are within $\pm$15\% of our model's predicted \hcfirst values, with a median error of 6.2\%.

We fit and validate our model on 24 commercial off-the-shelf DDR4 modules with DRAM chips from five different manufacturers and one HBM2 stack. For each module, we numerically fit our model to 72 \hcfirst measurements, and validate it against 178 new measurements on 1) new victim rows, and 2) 53 new access patterns and timings, including timings beyond the fitted range. Our results show that:
1)~94.4\% of our model's predicted \hcfirst values agree with the validation measurements ($\pm$15\%), with a median error of 5.4\%;
2)~our model is as accurate on new victim rows (95.0\%) as at new access patterns and timings (94.7\%), when both are new (93.4\%), and beyond the fitted range (95.9\%); and
3)~our model is equally accurate on the HBM2 stack (93.8\%).

We present two use cases that show how architects can use our model to better evaluate and mitigate a system's vulnerability to DRAM read disturbance. First, we show that by integrating our model into gem5 and Ramulator 2, we can accurately reproduce RowPress bitflips in full-system simulation using the user-space proof-of-concept RowPress program. Second, we demonstrate how our model can guide the design and configuration of DRAM read disturbance mitigations to more accurately account for RowPress (i.e., the reduction in \hcfirst as \tAggON increases) compared to prior works.

We make the following key contributions in this paper:
\begin{itemize}
\item To our knowledge, this is the first work to build an architecture-level DRAM read disturbance error model principled on device-level physical mechanisms of DRAM read disturbance.
\item We validate our model on 24 DDR4 modules from five manufacturers and one HBM2 stack and show that the model accurately predicts both 1) the numerical values of \hcfirst, and 2) how \hcfirst changes as the DRAM access pattern, timing, and data pattern change.
\item We demonstrate how architects can use our model to better evaluate and mitigate a system's vulnerability to DRAM read disturbance.
\item We will open-source our model.
\end{itemize}

\section{Background}
\label{sec:background}

\begin{table*}[t]
\caption{Summary of device-level mechanisms of DRAM read disturbance.}
\label{tab:bg-mechanisms}
\centering
\footnotesize
\renewcommand{\arraystretch}{1.1}
\begin{tabular}{@{}>{\raggedright\arraybackslash}p{0.09\textwidth}>{\raggedright\arraybackslash}p{0.10\textwidth}>{\raggedright\arraybackslash}p{0.21\textwidth}>{\raggedright\arraybackslash}p{0.53\textwidth}@{}}
\toprule
\textbf{Phenomenon} & \textbf{Bitflip Direction} & \textbf{Mechanism} & \textbf{Device-Level Process} \\
\midrule
RowHammer &
1-to-0 &
NWL-side electron migration~\cite{ryu2017overcoming, yang2019trap, walker2021ondramrowhammer, zhou2023double} &
While the \NWL is open, interface traps near it capture electrons. After it closes, the traps emit the electrons, and some migrate into the victim storage node. \\
RowHammer &
0-to-1 &
PWL-side electron migration~\cite{Gautam2020MitigatingPassing, zhou2023double} &
While the \PWL is open, its electric field pulls electrons out of the victim storage node. After it closes, not all of them return. \\
RowHammer &
Both &
Double-sided enhancement~\cite{zhou2023double, luo2026demystifying} &
While the traps near the closed aggressor emit, the open aggressor pulls the emitted electrons toward (\NWL closed, \PWL open) or away from (\PWL closed, \NWL open) the victim. \\
RowHammer &
Both &
Capacitive crosstalk~\cite{Zhou2024Unveiling, Zhou2024Understanding, Jie2024Understanding, luo2026demystifying, bai2026unified} &
Aggressor wordline edges couple into the victim wordline through parasitic capacitance and make the victim's access transistor leakier during the transient. \\
\midrule
RowPress &
0-to-1 &
NWL-induced leakage~\cite{Zhou2024Understanding, Zhou2024Unveiling} &
A long-open \NWL weakens the gate control of the victim's access transistor and draws electrons out of the storage node. Later works find this leakage too weak to cause bitflips in real chips~\cite{luo2026demystifying} and report that electrons accumulate in the storage node under a sustained \NWL bias~\cite{bai2026unified}. \\
RowPress &
1-to-0 &
PWL-induced leakage~\cite{Zhou2024Understanding, Zhou2024Unveiling} &
A long-open \PWL draws electrons from the substrate, and some are injected into the victim storage node. \\
\midrule
Both &
Both &
Bitline bias (sense-amplifier-mediated disturb)~\cite{bai2026unified} &
An activation lets the sense amplifier drive the shared bitline to the aggressor data's full level, while the victim wordline stays off. The bitline--storage-node voltage difference sets the victim's off-state leakage for as long as the aggressor row is open. \\
Both &
Both &
Tail and bulk cells~\cite{luo2026demystifying, bai2026unified} &
A few cells with an efficient trap fail first. Most cells lack such a trap and fail only at higher activation counts, through crosstalk and bitline bias. Manufacturing forms more traps near the \NWL than near the \PWL. \\
\bottomrule
\end{tabular}
\end{table*}

\subsection{DRAM Organization and Operation}
\label{sec:bg-organization}

\begin{figure}[t]
\centering
\definecolor{layInk}{HTML}{2B2F36}
\definecolor{layIso}{HTML}{F3F3F1}
\definecolor{layEdge}{HTML}{ABABAB}
\definecolor{layActive}{HTML}{DCDCDA}
\definecolor{layActiveCut}{HTML}{C3C3C0}
\definecolor{layWL}{HTML}{7A8494}
\definecolor{layWLTop}{HTML}{939DAB}
\definecolor{layBL}{HTML}{2E3A48}
\definecolor{laySNC}{HTML}{5E8D7B}
\definecolor{laySNCEdge}{HTML}{456B5D}
\definecolor{layNWL}{HTML}{DB2828}
\definecolor{layPWL}{HTML}{2185D0}
\definecolor{layPlug}{HTML}{E9E9E7}
\definecolor{layCap}{HTML}{ECE6DA}
\begin{tikzpicture}[
    x=1cm, y=1cm,
    every node/.style={font=\fontsize{6}{7}\selectfont, inner sep=1pt, text=layInk},
    tiny/.style={font=\fontsize{5}{5.6}\selectfont\bfseries, align=center, inner sep=0pt},
    paneltitle/.style={font=\fontsize{6.4}{7.6}\selectfont\bfseries},
    edge/.style={draw=layEdge, line width=0.35pt},
    active/.style={fill=layActive, edge, rounded corners=0.4pt},
    snc/.style={fill=laySNC, draw=laySNCEdge, line width=0.35pt},
    blc/.style={fill=white, draw=layBL, line width=0.6pt},
  ]

  \begin{scope}
    \clip (0, 0) rectangle (2.9, 2.52);
    \fill[white] (0, 0) rectangle (2.9, 2.52);
    \foreach \j in {-1, ..., 5} {
      \foreach \g in {-1, ..., 7} {
        \pgfmathtruncatemacro{\odd}{mod(\j + \g + 100, 2)}
        \ifnum\odd=1
          \pgfmathsetmacro{\x}{0.31 + 0.58*\j}
          \pgfmathsetmacro{\y}{2.52 - 0.42*\g}
          \path[active] (\x + 0.153, \y + 0.553) -- ++(0.18, -0.086) -- ++(-0.486, -1.02) -- ++(-0.18, 0.086) -- cycle;
        \fi
      }
    }
    \path[active, fill=layActiveCut] (1.47 + 0.153, 1.26 + 0.553) -- ++(0.18, -0.086) -- ++(-0.486, -1.02) -- ++(-0.18, 0.086) -- cycle;
    \foreach \k in {0, ..., 5} \fill[layWLTop] (0, 2.31 - 0.42*\k - 0.05) rectangle (2.9, 2.31 - 0.42*\k + 0.05);
    \foreach \j in {0, ..., 4} \fill[layBL] (0.31 + 0.58*\j - 0.032, 0) rectangle (0.31 + 0.58*\j + 0.032, 2.52);
    \foreach \j in {-1, ..., 5} {
      \foreach \g in {-1, ..., 7} {
        \pgfmathtruncatemacro{\odd}{mod(\j + \g + 100, 2)}
        \ifnum\odd=1
          \pgfmathsetmacro{\x}{0.31 + 0.58*\j}
          \pgfmathsetmacro{\y}{2.52 - 0.42*\g}
          \path[snc] (\x + 0.2, \y + 0.42) circle (0.085);
          \path[snc] (\x - 0.2, \y - 0.42) circle (0.085);
          \path[blc] (\x, \y) circle (0.08);
        \fi
      }
    }
    \draw[layInk, line width=0.6pt, dash pattern=on 1.6pt off 1.1pt] (1.47 - 0.32, 1.26 - 0.67) -- (1.47 + 0.32, 1.26 + 0.67);
    \foreach \cx/\cy/\cell in {1.67/1.68/1, 1.27/0.84/2}
      \node[circle, snc, minimum size=0.17cm, inner sep=0pt, text=white,
            font=\fontsize{4.4}{4.4}\selectfont\bfseries] at (\cx, \cy) {\cell};
  \end{scope}
  \draw[layInk, line width=0.5pt] (0, 0) rectangle (2.9, 2.52);
  \foreach \y/\name in {1.89/PWL\textsubscript{1}, 1.47/WL\textsubscript{1}, 1.05/WL\textsubscript{2}, 0.63/PWL\textsubscript{2}}
    \node[anchor=west] at (2.96, \y) {\name};
  \fill[layBL] (0.035, -0.39) rectangle (0.085, -0.19);   \node[anchor=west] at (0.11, -0.29) {BL};
  \path[active] (0.62 + 0.035, -0.29 + 0.11) -- ++(0.09, -0.043) -- ++(-0.17, -0.18) -- ++(-0.09, 0.043) -- cycle;
  \node[anchor=west] at (0.75, -0.29) {active region};
  \path[snc] (2.0, -0.29) circle (0.075);   \node[anchor=west] at (2.09, -0.29) {SNC};
  \path[blc] (2.66, -0.29) circle (0.07);   \node[anchor=west] at (2.75, -0.29) {BLC};
  \node[paneltitle] at (1.45, -0.68) {(a) Top view of a 6F\textsuperscript{2} array};

  \begin{scope}[shift={(3.95, 0)}]
    \path[fill=layActive, edge] (0.40, 0) rectangle (3.66, 1.25);
    \foreach \left/\gate/\name/\fillcolor in {0.10/0.22/PWL\\1/layPWL, 3.30/3.42/PWL\\2/layWL} {
      \path[fill=layIso, edge] (\left, 1.25) -- (\left, 0.48) arc[start angle=180, end angle=360, radius=0.33] -- (\left + 0.66, 1.25);
      \fill[\fillcolor] (\gate, 1.0) -- (\gate, 0.67) arc[start angle=180, end angle=360, radius=0.21] -- (\gate + 0.42, 1.0) -- cycle;
      \node[tiny, text=white] at (\gate + 0.21, 0.73) {\name};
      \node[tiny, text=black!55, font=\fontsize{4.6}{5}\selectfont] at (\left + 0.33, 0.29) {STI};
    }
    \foreach \left/\name/\fillcolor in {1.24/WL\\1/layWL, 2.30/WL\\2/layNWL} {
      \path[fill=layIso, edge] (\left, 1.25) -- (\left, 0.52) arc[start angle=180, end angle=360, radius=0.26] -- (\left + 0.52, 1.25);
      \fill[\fillcolor] (\left + 0.05, 0.98) -- (\left + 0.05, 0.55) arc[start angle=180, end angle=360, radius=0.21] -- (\left + 0.47, 0.98) -- cycle;
      \fill[layCap] (\left + 0.05, 0.98) rectangle (\left + 0.47, 1.25);
      \node[tiny, text=white] at (\left + 0.26, 0.66) {\name};
    }
    \path[snc] (0.76, 0.80) rectangle (1.24, 1.25);   \node[tiny, text=white] at (1.00, 1.025) {SNC\\1};
    \path[snc] (2.82, 0.80) rectangle (3.30, 1.25);   \node[tiny, text=white] at (3.06, 1.025) {SNC\\2};
    \path[blc] (1.79, 0.84) rectangle (2.27, 1.22);   \node[tiny, text=layBL] at (2.03, 1.03) {BLC};
    \path[fill=layPlug, edge] (1.93, 1.22) rectangle (2.13, 1.54);
    \fill[layBL] (0.25, 1.54) rectangle (3.81, 1.72);
    \node[tiny, text=white] at (2.03, 1.63) {BL};
    \foreach \x/\label in {1.00/{cell 1 (victim)}, 3.06/{cell 2}} {
      \path[fill=layPlug, edge] (\x - 0.1, 1.25) rectangle (\x + 0.1, 1.96);
      \fill[layPlug] (\x - 0.26, 1.96) rectangle (\x + 0.26, 2.42);
      \fill[white] (\x - 0.14, 2.08) rectangle (\x + 0.14, 2.42);
      \draw[layEdge, line width=0.35pt] (\x - 0.26, 2.42) -- (\x - 0.26, 1.96) -- (\x + 0.26, 1.96) -- (\x + 0.26, 2.42)
                                       (\x - 0.14, 2.42) -- (\x - 0.14, 2.08) -- (\x + 0.14, 2.08) -- (\x + 0.14, 2.42);
      \node[font=\fontsize{6}{7}\selectfont\bfseries] at (\x, 2.56) {\label};
    }
    \node[text=layPWL, font=\fontsize{5.6}{6.4}\selectfont\bfseries] at (0.43, -0.2) {PWL of cell 1};
    \node[text=layNWL, font=\fontsize{5.6}{6.4}\selectfont\bfseries] at (2.56, -0.2) {NWL of cell 1};
    \node[paneltitle] at (1.98, -0.68) {(b) Cross-section along the dashed line};
  \end{scope}
\end{tikzpicture}
\caption{Physical layout of a $\mathrm{6F^2}$ DRAM cell array, reproduced from~\cite{luo2025revisiting}: (a) top view; (b) cross-section along the dashed line in (a), where two cells share one active region and one bitline contact. SNC: storage node contact; BLC: bitline contact; STI: shallow trench isolation.}
\label{fig:cell-layout}
\end{figure}
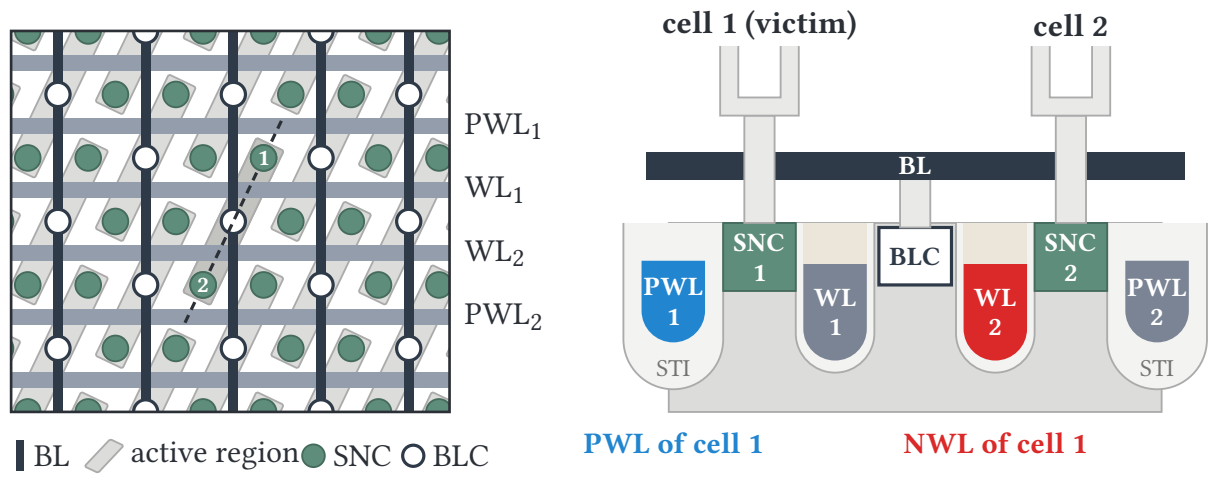

A DRAM module consists of chips that are organized hierarchically into banks, subarrays, and rows~\cite{kim2012case, kim2014flipping}. A DRAM cell stores one bit as charge in a capacitor that connects to a bitline through an access transistor, whose gate is a wordline (WL; Figure~\ref{fig:cell-layout}). Opening (i.e., \emph{activating}, ACT) a row raises its wordline, connects the row's cells to their bitlines, and lets the sense amplifiers latch the row's data. Closing (i.e., \emph{precharging}, PRE) the row lowers its wordline and returns the bitlines to their precharge voltage. We call the time a row stays open the \emph{aggressor on time} (\tAggON) and the time between closing it and the next activation the \emph{aggressor off time} (\tAggOFF)~\cite{luo2023rowpress}.

\head{Physical Cell Layout} The address a memory controller uses for a row (the \emph{logical} row) generally differs from the row's position in the array (the \emph{physical} row), so two logically consecutive rows need not be physically adjacent~\cite{kim2014flipping, kim2020revisiting, nam2024dramscope}. In the $\mathrm{6F^2}$ array layout of modern chips (Figure~\ref{fig:cell-layout}a), two cells share one active region and one bitline contact (Figure~\ref{fig:cell-layout}b). For a victim cell, the wordline of the other cell in its active region is the \emph{neighboring wordline} (\NWL; WL\textsubscript{2} for cell~1 in Figure~\ref{fig:cell-layout}b), and the wordline that only passes by the active region is the \emph{passing wordline} (\PWL; PWL\textsubscript{1})~\cite{yang2019trap, zhou2023double, luo2026demystifying}. Because cells and wordlines alternate in this layout, one aggressor row is the \NWL for half of the cells in a victim row and the \PWL for the other half~\cite{luo2026demystifying}.

\subsection{DRAM Read Disturbance}
\label{sec:bg-read-disturbance}

\emph{Read disturbance} is the phenomenon where accessing a DRAM row (the \emph{aggressor} row) disturbs the data stored in other, unaccessed rows (the \emph{victim} rows)~\cite{kim2014flipping, mutlu2019rowhammer, mutlu2023retrospective}. \head{RowHammer} Repeatedly activating and precharging an aggressor row many times causes bitflips in physically nearby victim rows~\cite{kim2014flipping, kim2020revisiting, orosa2021deeper}. \head{RowPress} Keeping an aggressor row open for a long time (i.e., a large \tAggON) causes bitflips with many fewer activations than RowHammer needs~\cite{luo2023rowpress, luo2024experimental}. A victim row can be disturbed from one side (\emph{single-sided}) or from both sides with alternating activations (\emph{double-sided})~\cite{kim2020revisiting, hassan2021utrr}.

\head{Read Disturbance Metric} We use \hcfirst, the minimum aggressor row activation count to induce at least one bitflip, counting the activations of both aggressor rows~\cite{kim2020revisiting, yaglikci2024spatial}. \hcfirst is the quantity that read-disturbance mitigations are configured with~\cite{yaglikci2021blockhammer, park2020graphene, qureshi2022hydra, olgun2024abacus, canpolat2025chronus}. It depends on the access pattern, on the data stored in the victim and aggressor rows, on the module~\cite{kim2020revisiting, orosa2021deeper, luo2023rowpress, yaglikci2024spatial}, and it varies over time and across rows~\cite{olgun2025variable}. We refer to a bitflip by its direction, \emph{1-to-0} or \emph{0-to-1}, and to a data pattern by the charge state of the victim and aggressor cells (e.g., \vzero: victim cells discharged, aggressor cells charged).

\subsection{Device-Level Modeling of Read Disturbance}
\label{sec:bg-mechanisms}

Device-level works use technology computer-aided design (TCAD) simulations and wafer-level measurements to explain read-disturbance bitflips~\cite{ryu2017overcoming, yang2019trap, walker2021ondramrowhammer, Gautam2020MitigatingPassing, zhou2023double, Zhou2024Unveiling, Zhou2024Understanding, park2014active, park2016experiments, yang2016suppression, gautam2019rowhammering, han2021surround}. Prior works~\cite{luo2026demystifying, bai2026unified} reconcile these explanations with experimental characterization and summarize the first-order mechanisms we build on. First, interface traps near the \NWL or the \PWL capture electrons while that wordline is high and emit them after it falls; electrons from \NWL-side traps migrate \emph{into} the victim's storage node (1-to-0 bitflips), while \PWL-side traps draw electrons \emph{out of} it (0-to-1 bitflips)~\cite{yang2019trap, zhou2023double, luo2026demystifying}. Second, the trap location and the electron-migration path set the direction and the efficiency of each mechanism~\cite{luo2026demystifying, bai2026unified}. Third, a long-open adjacent row causes additional leakage of the victim, which explains RowPress~\cite{Zhou2024Understanding, Zhou2024Unveiling, luo2026demystifying}; this leakage is driven mainly by the bitline voltage that the sense amplifier holds while the adjacent row is open~\cite{bai2026unified}. Table~\ref{tab:bg-mechanisms} summarizes these mechanisms with their bitflip directions and key dependences. These works explain \emph{why} bitflips occur, but a TCAD deck of a specific commodity chip is not available, and a TCAD simulation cannot be fitted to, or run alongside, the access stream of an architecture simulator.

\section{Motivation and Related Work}
\label{sec:motivation}
\label{sec:related}

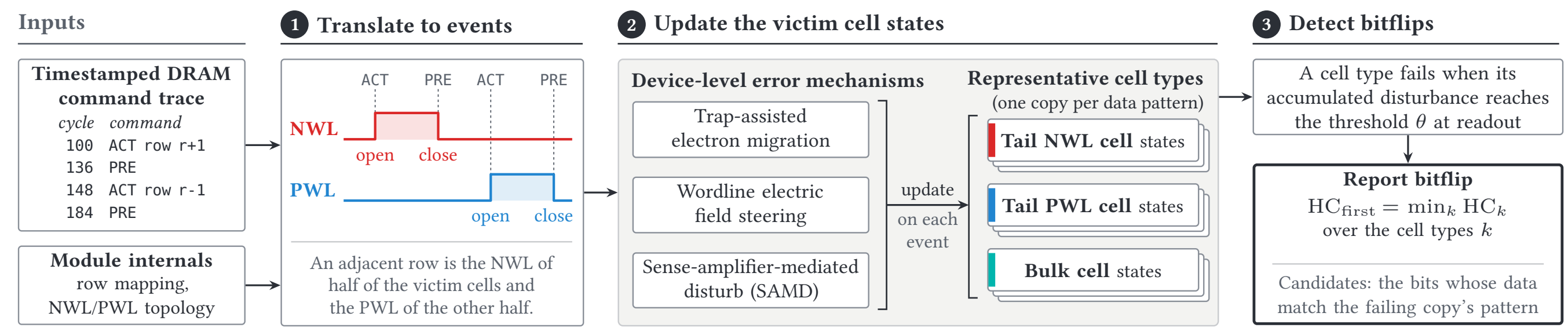
\begin{figure*}[t]
\centering
\definecolor{pipeInk}{HTML}{2B2F36}
\definecolor{pipeEdge}{HTML}{8A9099}
\definecolor{pipeMuted}{HTML}{5A616B}
\definecolor{pipePanel}{HTML}{F3F3F1}
\definecolor{pipeRule}{HTML}{C9CBCE}
\definecolor{pipeNWL}{HTML}{DB2828}
\definecolor{pipePWL}{HTML}{2185D0}
\definecolor{pipeBulk}{HTML}{00B5AD}
\providecommand{\pipeFirstMechanism}{Trap-assisted\\electron migration}
\begin{tikzpicture}[
    x=1cm, y=1cm,
    every node/.style={font=\fontsize{6.6}{7.9}\selectfont, inner sep=2pt, text=pipeInk},
    note/.style={font=\fontsize{6.2}{7.3}\selectfont},
    code/.style={font=\fontsize{6}{7.3}\ttfamily\selectfont},
    grouplabel/.style={font=\fontsize{6.8}{8}\selectfont\bfseries, inner sep=0pt},
    stagetitle/.style={font=\fontsize{7.4}{8.8}\selectfont\bfseries, anchor=west, inner sep=0pt},
    stepmark/.style={circle, fill=pipeInk, text=white, font=\fontsize{6.2}{7}\selectfont\bfseries, inner sep=0pt,
                     minimum size=3.2mm},
    box/.style={draw=pipeEdge, line width=0.45pt, rounded corners=1pt, fill=white, align=center},
    panel/.style={draw=pipeRule, line width=0.45pt, rounded corners=1.5pt, fill=pipePanel},
    celltype/.style={box, text width=2.3cm, minimum height=0.48cm, inner sep=1.5pt,
                     double copy shadow={shadow xshift=1.7pt, shadow yshift=-1.7pt, draw=pipeEdge, fill=white}},
    headrule/.style={draw=pipeInk, line width=0.5pt},
    flow/.style={-{Stealth[length=3.6pt, width=3pt]}, line width=0.6pt, draw=pipeInk},
    bracket/.style={draw=pipeInk, line width=0.5pt},
    guide/.style={draw=pipeMuted, line width=0.35pt, dash pattern=on 0.8pt off 0.9pt},
  ]

  \def\xInputL{0}       \def\xInputR{2.6}
  \def\xEventsL{3.0}    \def\xEventsR{6.45}
  \def\xStateL{6.85}    \def\xStateR{13.70}
  \def\xDetectL{14.10}  \def\xDetectR{17.60}
  \def\yHead{3.43}
  \def\yRule{3.21}
  \def\yTop{3.04}
  \def\yBottom{0}

  \def\pipeStage#1#2#3#4{%
    \node[stepmark] at (#1 + 0.16, \yHead) {#3};
    \node[stagetitle] at (#1 + 0.41, \yHead) {#4};
    \draw[headrule] (#1, \yRule) -- (#2, \yRule);
  }
  \node[stagetitle, text=pipeMuted] at (\xInputL, \yHead) {Inputs};
  \draw[headrule, draw=pipeMuted] (\xInputL, \yRule) -- (\xInputR, \yRule);
  \pipeStage{\xEventsL}{\xEventsR}{1}{Translate to events}
  \pipeStage{\xStateL}{\xStateR}{2}{Update the victim cell states}
  \pipeStage{\xDetectL}{\xDetectR}{3}{Detect bitflips}

  \node[box, anchor=north west, text width=2.6cm - 4pt - 0.45pt] (trace) at (\xInputL, \yTop)
    {\textbf{Timestamped DRAM}\\\textbf{command trace}\\[1.5pt]
     {\fontsize{6}{7.3}\ttfamily\selectfont
      \begin{tabular}{@{}r@{\hspace{5pt}}l@{}}
        {\rmfamily\itshape cycle} & {\rmfamily\itshape command}\\
        100 & ACT row r+1\\
        136 & PRE\\
        148 & ACT row r-1\\
        184 & PRE
      \end{tabular}}};
  \node[box, anchor=south west, text width=2.6cm - 4pt - 0.45pt] (internals) at (\xInputL, \yBottom)
    {\textbf{Module internals}\\{\fontsize{6.2}{7.3}\selectfont row mapping,\\NWL/PWL topology}};

  \node[box, anchor=north west, minimum width=3.45cm, minimum height=3.04cm] (events) at (\xEventsL, \yTop) {};
  \draw[flow] (trace.east) -- (trace.east -| events.west);
  \draw[flow] (internals.east) -- (internals.east -| events.west);

  \def\xWave{3.72}
  \def\xWaveEnd{6.33}
  \def\yCommands{2.80}
  \def\yNWL{2.12}
  \def\yPWL{1.42}
  \def\pulseHeight{0.30}
  \def\xNWLopen{4.08}  \def\xNWLclose{4.80}  \def\xPWLopen{5.40}  \def\xPWLclose{6.12}

  \foreach \x/\command/\yEdge in {\xNWLopen/ACT/\yNWL, \xNWLclose/PRE/\yNWL, \xPWLopen/ACT/\yPWL, \xPWLclose/PRE/\yPWL} {
    \node[code, text=pipeMuted, inner sep=1pt] (tag) at (\x, \yCommands) {\command};
    \draw[guide] (tag.south) -- (\x, \yEdge + \pulseHeight);
  }
  \node[anchor=west, inner sep=0pt, font=\fontsize{6.6}{7.9}\selectfont\bfseries, text=pipeNWL] at (3.1, \yNWL + 0.12) {NWL};
  \node[anchor=west, inner sep=0pt, font=\fontsize{6.6}{7.9}\selectfont\bfseries, text=pipePWL] at (3.1, \yPWL + 0.12) {PWL};
  \fill[pipeNWL!14] (\xNWLopen, \yNWL) rectangle (\xNWLclose, \yNWL + \pulseHeight);
  \fill[pipePWL!14] (\xPWLopen, \yPWL) rectangle (\xPWLclose, \yPWL + \pulseHeight);
  \draw[pipeNWL, line width=0.8pt, line join=round] (\xWave, \yNWL) -- (\xNWLopen, \yNWL) -- ++(0, \pulseHeight)
      -- (\xNWLclose, \yNWL + \pulseHeight) -- (\xNWLclose, \yNWL) -- (\xWaveEnd, \yNWL);
  \draw[pipePWL, line width=0.8pt, line join=round] (\xWave, \yPWL) -- (\xPWLopen, \yPWL) -- ++(0, \pulseHeight)
      -- (\xPWLclose, \yPWL + \pulseHeight) -- (\xPWLclose, \yPWL) -- (\xWaveEnd, \yPWL);
  \foreach \x/\event/\y/\tint in {\xNWLopen/open/\yNWL/pipeNWL, \xNWLclose/close/\yNWL/pipeNWL,
                                   \xPWLopen/open/\yPWL/pipePWL, \xPWLclose/close/\yPWL/pipePWL}
    \node[note, text=\tint, inner sep=0pt, anchor=base] at (\x, \y - 0.24) {\event};
  \draw[draw=pipeRule, line width=0.45pt] (\xEventsL + 0.12, 0.95) -- (\xEventsR - 0.12, 0.95);
  \node[note, text=pipeMuted, align=center, anchor=south, inner sep=0pt] at (4.725, 0.12)
    {An adjacent row is the NWL of\\half of the victim cells and\\the PWL of the other half.};

  \def\yPanelTop{3.04}    \def\yPanelBottom{0}
  \node[panel, anchor=north west, minimum width=6.85cm, minimum height=\yPanelTop cm - \yPanelBottom cm] (state)
    at (\xStateL, \yPanelTop) {};
  \draw[flow] (events.east) -- (events.east -| state.west);

  \def\xMech{8.37}
  \def\xCells{12.28}
  \def\yGroupLabel{2.80}
  \node[grouplabel, anchor=west] at (7.0, \yGroupLabel) {Device-level error mechanisms};
  \node[grouplabel, anchor=east] at (13.55, \yGroupLabel) {Representative cell types};
  \node[note, anchor=east, inner sep=0pt] at (13.55, 2.52) {(one copy per data pattern)};

  \node[box, text width=2.55cm, minimum height=0.58cm] (mechFirst) at (\xMech, 2.23) {\pipeFirstMechanism};
  \node[box, text width=2.55cm, minimum height=0.58cm] (mechSteer) at (\xMech, 1.37) {Wordline electric\\field steering};
  \node[box, text width=2.55cm, minimum height=0.58cm] (mechLast) at (\xMech, 0.51)
    {Sense-amplifier-mediated\\disturb (SAMD)};

  \foreach \y/\name/\tint/\i in {2.11/Tail NWL cell/pipeNWL/1, 1.37/Tail PWL cell/pipePWL/2, 0.63/Bulk cell/pipeBulk/3} {
    \node[celltype] (cell\i) at (\xCells, \y) {\textbf{\name} {\fontsize{6.2}{7.3}\selectfont states}};
    \draw[\tint, line width=2.2pt] ([xshift=1.5pt, yshift=1.6pt]cell\i.south west) -- ([xshift=1.5pt, yshift=-1.6pt]cell\i.north west);
  }

  \def\xBracketMech{9.92}    \def\xBracketCells{10.86}
  \draw[bracket] (\xBracketMech - 0.1, 2.56) -- (\xBracketMech, 2.56) -- (\xBracketMech, 0.18) -- (\xBracketMech - 0.1, 0.18);
  \draw[bracket] (\xBracketCells + 0.1, 2.39) -- (\xBracketCells, 2.39) -- (\xBracketCells, 0.31) -- (\xBracketCells + 0.1, 0.31);
  \draw[flow] (\xBracketMech, 1.37) -- (\xBracketCells, 1.37);
  \node[note, anchor=south, inner sep=1.5pt] at (10.39, 1.37) {update};
  \node[note, anchor=north, inner sep=2pt, text=pipeMuted, align=center] at (10.39, 1.37) {on each\\event};

  \node[box, anchor=north west, text width=3.55cm - 4pt - 0.45pt] (check) at (\xDetectL, \yTop)
    {A cell type fails when its\\accumulated disturbance reaches\\the threshold $\theta$ at readout};
  \node[box, line width=0.9pt, draw=pipeInk, anchor=south west, text width=3.55cm - 4pt - 0.9pt] (report) at (\xDetectL, \yBottom)
    {\textbf{Report bitflip}\\[1pt]$\mathrm{HC}_{\mathrm{first}}\,{=}\,\min_{k} \mathrm{HC}_{k}$\\[-0.5pt]
     {\fontsize{6.2}{7.3}\selectfont over the cell types $k$}\\[1.5pt]
     {\color{pipeRule}\rule{3.1cm}{0.45pt}}\\[0.5pt]
     \fontsize{6.2}{7.3}\selectfont\color{pipeMuted}Candidates:\ the bits whose data\\match the failing copy's pattern};
  \draw[flow] (state.east |- check.west) -- (check.west);
  \draw[flow] (check.south) -- (report.north);

\end{tikzpicture}
\caption[Overview of the model]{Overview of \model. \protect\circled{1}~A timestamped DRAM command trace is translated, using the row mapping and the \NWL/\PWL topology of the module, into open and close events of the \NWL and the \PWL of a victim cell. \protect\circled{2}~On each event, the modeled device-level error mechanisms update the states of three representative cell types; \model keeps one copy of the states per data pattern. \protect\circled{3}~A cell type fails when its accumulated disturbance reaches the failure threshold $\theta$ at readout; the earliest cell type sets \hcfirst.}
\label{fig:model-pipeline}
\end{figure*}

\begin{table}[t]
\centering
\caption{The scope of prior modeling works of DRAM read disturbance.}
\label{tab:related-models}
\footnotesize
\setlength{\tabcolsep}{1.3pt}
\renewcommand{\arraystretch}{1.15}
\resizebox{\columnwidth}{!}{%
\begin{tabular}{@{}l l ccccc@{}}
\toprule
\textbf{Model} & \tabhead{\textbf{Abstraction}\\\textbf{level}}
 & \tabhead{\textbf{Physical}\\\textbf{mechanisms}}
 & \tabhead{\textbf{Access}\\\textbf{pattern}}
 & \tabhead{\textbf{Timing}\\\textbf{(\tAggON)}}
 & \tabhead{\textbf{Timing}\\\textbf{(\tAggOFF)}}
 & \tabhead{\textbf{Data}\\\textbf{pattern}} \\
\midrule
France et al.~\cite{france2021implementing, france2022modeling} & Architecture & \xmark & \xmark & \xmark & \xmark & \xmark \\
Hammulator~\cite{thomas2023hammulator} & Architecture & \xmark & \xmark & \xmark & \xmark & \xmark \\
HammerSim~\cite{goswami2026hammersimsystemleveltoolmodel} & Architecture & \xmark & \pmark & \xmark & \xmark & \xmark \\
DSAC~\cite{hong2023dsac} & Mitigation & \xmark & \xmark & \cmark & \xmark & \xmark \\
ImPress~\cite{qureshi2024impress} & Mitigation & \xmark & \xmark & \cmark & \xmark & \xmark \\
Demirel et al.~\cite{demirel2025circuit} & Circuit (SPICE) & \cmark & \pmark & \pmark & \cmark & \xmark \\
Jiang et al.~\cite{jiang2021quantifying} & Circuit (RC) & \pmark & \xmark & \xmark & \xmark & \xmark \\
Hu et al.~\cite{hu2026quantifying} & Circuit (RC) & \pmark & \xmark & \pmark & \xmark & \pmark \\
\midrule
\textbf{\Model} & Architecture & \cmark & \cmark & \cmark & \cmark & \cmark \\
\bottomrule
\end{tabular}}
\end{table}

Accurately modeling DRAM read disturbance at the architecture level is important for 1) evaluating a system's vulnerability to DRAM read disturbance bitflips~\cite{goswami2026hammersimsystemleveltoolmodel, thomas2023hammulator} for security research, and 2) designing and configuring DRAM read disturbance mitigation techniques to account for different DRAM access patterns and timings~\cite{luo2023rowpress,hong2023dsac,qureshi2024impress}. Unfortunately, existing architecture-level DRAM read disturbance models do \emph{not} fully consider how DRAM access patterns, timings, and data patterns sophisticate DRAM read disturbance bitflips (Table~\ref{tab:related-models}). Table~\ref{tab:related-models} compares the modeling scope of existing DRAM read disturbance models in prior works.

\head{Existing Architecture-Level Models} Hammulator~\cite{thomas2023hammulator}, HammerSim~\cite{goswami2026hammersimsystemleveltoolmodel}, and France et al.~\cite{france2021implementing, france2022modeling} model read-disturbance bitflips in architecture simulators. Hammulator and France et al. simply count the DRAM row activations within a refresh window and trigger bitflips once the count exceeds a pre-determined \hcfirst; HammerSim also distinguishes single-sided, double-sided, and Half-Double~\cite{kogler2022half} access patterns, each with its own bitflip probability. These models do \emph{not} consider the DRAM timings and data patterns.

\head{RowPress Models from Mitigations} DSAC~\cite{hong2023dsac} and ImPress~\cite{qureshi2024impress} model RowPress by scaling the hammer count with \tAggON logarithmically (DSAC) or linearly (ImPress) to adapt RowHammer mitigations to also mitigate RowPress. However, they do not distinguish access patterns, and do not model \tAggOFF and data patterns.

\head{Circuit-Level Models} Demirel et al.~\cite{demirel2025circuit} build a SPICE circuit model of a small DRAM cell array mainly to model \tAggOFF. In their model, each toggle of an \NWL aggressor injects charge into the victim cell with an exponentially decaying leakage, and an open \PWL aggressor discharges the victim cell at a constant rate. However, they do not fully model \tAggON with different DRAM access patterns and different data patterns. Jiang et al.~\cite{jiang2021quantifying} and Hu et al.~\cite{hu2026quantifying} model a DRAM cell as an RC circuit with two leakage paths. However, they evaluate only double-sided access patterns at one or two aggressor on times, with at most two data patterns.

\section{Our Modeling Methodology}
\label{sec:model}

Figure~\ref{fig:model-pipeline} shows the high-level workflow of our architecture-level error model of DRAM read disturbance in three steps.

\head{\circled{1} Translating Commands to Events}\label{sec:model-state} Our model takes a timestamped DRAM command trace (i.e., the ACT and PRE commands with addresses and clock cycles, from, e.g., Ramulator~2~\cite{luo2023ramulator2}) as input. The model then translates the traces into four aggressor row events that disturb the DRAM cells in the adjacent victim row(s): \texttt{NWL open}, \texttt{NWL close}, \texttt{PWL open}, and \texttt{PWL close}, using the row mapping and the \NWL/\PWL topology of the DRAM array. Because an adjacent row is the \NWL for half of the cells in a victim row and the \PWL for the other half (Section~\ref{sec:bg-organization}), each activation and precharge of an adjacent row becomes an \NWL event for one cell orientation and a \PWL event for the other.

\head{\circled{2} Updating the State of Victim Cell}\label{sec:model-cell-types} Upon a new event, our model updates the states (e.g., the accumulated disturbance $D$) of the disturbed victim cells using the modeled device-level error mechanisms (Section~\ref{sec:model-mechanisms}).  For a DRAM row, our model abstracts it into representative and distinct types of victim cells instead of tracking every single individual DRAM cell, to avoid intractable storage overhead. Based on device-level works~\cite{luo2026demystifying, bai2026unified}, we identify three types of victim cells: 1)~\emph{tail \NWL} cells, which have an efficient \NWL-side trap and cause the first 1-to-0 RowHammer bitflips, 2)~\emph{tail \PWL} cells, which have an efficient \PWL-side trap and cause the first 0-to-1 RowHammer bitflips, and 3)~\emph{bulk} cells, which have weak traps on both sides, and cause RowPress bitflips at a long \tAggON.

\head{\circled{3} Detecting Bitflips}\label{sec:model-hcfirst} Our model identifies read disturbance bitflips by checking whether the accumulated disturbance $D$ of any cell type reaches a failure threshold $\theta$. A victim cell that stores 0 fails when $D \geq \theta$ at readout, and one that stores 1 fails when $-D \geq \theta$; a transient crossing before readout is not a failure. Because our model works at the row level, it keeps one state for each local data pattern, i.e., the charge states of a victim cell and of its \NWL and \PWL aggressor cells. When the state of a data pattern reaches $\theta$, the bits of the victim row whose current data match that pattern are the bitflip candidates.

\subsection{Victim Cell States}
\label{sec:model-cell-state}

For each victim cell type of a (victim) DRAM row, \model maintains three kinds of states.

\head{Accumulated Disturbance} Our model keeps track of the accumulated disturbance $D$ to the victim cell voltage caused by aggressor wordline events (i.e., the opening and closing of the NWL and PWL). A positive $D$ will lead to 0-to-1 bitflips, and a negative $D$ will lead to 1-to-0 bitflips.

\head{Charge Trap Occupancies} Our model keeps track of the occupancies $o_{\rho,i}$ of the cell type's \NWL-side and \PWL-side charge traps ($\rho$ = \NWL or \PWL). To account for the variation in how fast charge traps capture and emit electrons, our model maintains multiple effective traps ($i = 1, \ldots, 12$) on each side, each with its own capture and emission rates. The occupancy $o_{\rho,i}$ increases as the traps capture electrons while the corresponding aggressor wordline is open, and decreases as the traps emit these electrons after the wordline closes.

\head{Wordline Event Timestamps} When a wordline is opened/closed, our model records the corresponding timestamps to determine how long the wordline has been open/closed that the modeled device-level mechanisms use as inputs.

\subsection{Modeled Physical Mechanisms}
\label{sec:model-mechanisms}

\head{Trap-Assisted Electron Migration} Charge traps near an aggressor wordline capture electrons while the wordline is open and emit them after it closes; the emitted electrons that reach the victim disturb its stored charge~\cite{yang2019trap, zhou2023double, luo2026demystifying}. To account for the variation in how fast charge traps capture and emit electrons, \model represents the traps near the adjacent wordline of type $\rho$ ($\rho$ = \NWL or \PWL) with twelve effective traps of equal capacity ($i = 1, \ldots, 12$). Because different traps capture and emit electrons at different rates, a single capture/emission rate cannot reproduce \hcfirst as \tAggON and \tAggOFF vary over two orders of magnitude. \Model therefore spreads the capture time constants $\tau_i$ of the effective traps log-uniformly between a fastest time $\tau_{\min}$ and a slowest time $10^{\Delta}\tau_{\min}$. While the wordline of type $\rho$ is open for time $t$, each effective trap captures electrons at rate $1/\tau_i$:
\begin{equation}
o_{\rho,i} \leftarrow o_{\rho,i} + (1-o_{\rho,i})\left(1-e^{-t/\tau_i}\right).
\label{eq:capture}
\end{equation}
After the wordline closes, each effective trap emits its electrons at the slower rate $1/(\epsilon\tau_i)$ and releases
\begin{equation}
r_{\rho,i} = o_{\rho,i}\left(1-e^{-t/(\epsilon\tau_i)}\right)
\label{eq:emission}
\end{equation}
over the wordline closed time $t$, where $\epsilon > 1$ is the emission-to-capture ratio. The charge released by the twelve effective traps of one side is summed (scaled by the \emph{trap strength} of the cell type on that side) and, after wordline electric field steering (described below), added to the accumulated disturbance $D$ in units of the \emph{disturbance scale}~$s$, a fraction of $\theta$ that sets the absolute \hcfirst. Released charge from \NWL-side traps moves electrons into the victim's storage node and drives it toward 1-to-0, and released charge from \PWL-side traps drives it toward 0-to-1~\cite{yang2019trap, zhou2023double, luo2026demystifying}. A tail \NWL cell has only \NWL-side traps (of strength~1), a tail \PWL cell has only \PWL-side traps (of strength $A_{\mathrm{PWL}}$), and a bulk cell has traps on both sides that are $L$ times as strong as those of the tail cells (i.e., $L$ on the \NWL side and $L A_{\mathrm{PWL}}$ on the \PWL side). Fast traps fill within a short \tAggON and empty within a short \tAggOFF, whereas slow traps fill only over a long or repeated \tAggON and keep their charge over many activations. Together, they explain why \hcfirst falls gradually as \tAggON grows (more capture) and rises as \tAggOFF shrinks (less emission before the next activation). We empirically model how the bitline voltage, which depends on the data stored in the aggressor cell, affects electron migration by scaling the released charge by
\begin{equation}
1 + \gamma \cdot c,
\label{eq:data-gain}
\end{equation}
where $c$ is $+1$ or $-1$ when the aggressor and victim data favor or oppose that side's bitflip direction and $0$ when they match, with separate fitted gains for each side's charge released while the bitline is precharged ($\gamma^{p}$) and while the other adjacent row is open and holds the bitline ($\gamma^{d}$).

\head{Wordline Electric Field Steering} The wordline's electric field when a wordline is open affects electron migration in the active region of the DRAM cell. When an aggressor wordline is closed (i.e., the charge traps near it start to emit electrons) while the other adjacent wordline is open, the electric field of the opened wordline steers the released electrons (toward the victim's storage node for \NWL-side traps and away from it for \PWL-side traps), resulting in a larger disturbance in the victim storage node. We model how the wordline electric field steers electrons with the field factor
\begin{equation}
F_\rho = T^{d}_{\rho}/T^{p}
\label{eq:field-factor}
\end{equation}
where $T^{d}_{\rho}$ and $T^{p}$ are the fractions of the released charge that disturb the victim with and without the other adjacent wordline open; $F_\rho$ states how many times more charge disturbs the victim when that wordline is open. Wordline electric field steering is why double-sided patterns, in which one aggressor wordline is open while the traps near the other emit, need much fewer aggressor row activations than single-sided ones.

\begin{table}[t]
\centering
\caption{Parameters of \model.}
\label{tab:parameters}
\footnotesize
\setlength{\tabcolsep}{3pt}
\renewcommand{\arraystretch}{1.08}
\begin{tabular}{@{}>{\raggedright\arraybackslash}p{0.19\columnwidth}l>{\raggedright\arraybackslash}p{0.57\columnwidth}@{}}
\toprule
\textbf{Mechanism} & \textbf{Symbol} & \textbf{Meaning} \\
\midrule
Disturbance scale & $s$ & change in accumulated disturbance ($D$) from a full release of traps of strength~1 while the other adjacent wordline is closed ($c = 0$), as a fraction of $\theta$ (sets the absolute \hcfirst) \\
\midrule
Trap-assisted & $\tau_{\min}$ & fastest capture time (ns) \\
electron migration & $\Delta$ & span of capture times (slowest: $10^{\Delta}\tau_{\min}$) \\
 & $\epsilon$ & emission-to-capture time ratio \\
\midrule
Trap strength & $A_{\mathrm{PWL}}$ & strength of the tail \PWL cell's \PWL-side traps relative to the tail \NWL cell's \NWL-side traps \\
 & $L$ & strength of the bulk cell's traps relative to the tail cells' traps \\
\midrule
Wordline electric & $F_{\mathrm{NWL}}$ & field factor of \NWL-side charge released while the \PWL is open \\
field steering & $F_{\mathrm{PWL}}$ & field factor of \PWL-side charge released while the \NWL is open \\
\midrule
Bitline data gain & $\gamma^{p}_{\mathrm{NWL}}$ & gain of \NWL-side charge released while the bitline is precharged \\
 & $\gamma^{d}_{\mathrm{NWL}}$ & gain of \NWL-side charge released while the \PWL holds the bitline \\
 & $\gamma^{p}_{\mathrm{PWL}}$ & gain of \PWL-side charge released while the bitline is precharged \\
 & $\gamma^{d}_{\mathrm{PWL}}$ & gain of \PWL-side charge released while the \NWL holds the bitline \\
\midrule
SAMD & $m_{\mathrm{tail},\mathrm{NWL}}$ & leakage rate of the tail cells while the \NWL is open ($s$/ms) \\
 & $m_{\mathrm{bulk},\mathrm{NWL}}$ & leakage rate of the bulk cell while the \NWL is open ($s$/ms) \\
 & $m_{\mathrm{bulk},\mathrm{PWL}}$ & leakage rate of the bulk cell while the \PWL is open ($s$/ms) \\
\bottomrule
\end{tabular}
\end{table}

\head{Sense-Amplifier-Mediated Disturb (SAMD)} The bitline voltage that the sense amplifier holds while a row is open affects the leakage of the other cells (i.e., the victim cell) that share the same bitlines. When an aggressor wordline is open and its row holds data opposite to the victim's, the sense amplifier drives the shared bitline to the full voltage level of the aggressor's data and holds it across the victim's OFF access transistor for as long as the aggressor row stays open~\cite{bai2026unified}, resulting in electrons leaking into the victim storage node~\cite{Zhou2024Understanding, Zhou2024Unveiling, luo2026demystifying}. We model this leakage as a constant drift of the accumulated disturbance toward 1-to-0,
\begin{equation}
\frac{dD}{dt} = -m_{k,\rho},
\label{eq:samd}
\end{equation}
where $m_{k,\rho}$ is the leakage rate of cell type $k$ while the adjacent wordline of type $\rho$ is open with opposite data. SAMD grows with \tAggON and is why \hcfirst of 1-to-0 bitflips falls steeply at a long \tAggON (i.e., RowPress).

\subsection{Model Parameters and Numerical Fitting}
\label{sec:model-fitting}

Table~\ref{tab:parameters} summarizes the parameters of \model. To determine the numerical values of the parameters, we fit them based on real-chip measurements of \hcfirst from commercial off-the-shelf DRAM modules. The model structure itself (e.g., the modeled device-level mechanisms, the three representative victim cell types) stays fixed and is not fitted or tuned for each module.

We search for the parameter values that minimize the sum of squared residuals $r$ over measured \hcfirst with different access patterns, timings, and data patterns of a module. We define the residual $r$ as the relative error of the predicted \hcfirst on a logarithmic scale (due to the vastly different \hcfirst values at different \tAggON and \tAggOFF), weighted by the uncertainty $\sigma$ of the measured value (Section~\ref{sec:metric}):
\begin{equation}
r = \frac{\ln(\mathrm{HC}_{\mathrm{model}} / \mathrm{HC}_{\mathrm{measured}})}{\sigma}.
\label{eq:residual}
\end{equation}

\section{Real-Chip Fitting and Validation Methodology}
\label{sec:methodology}

\subsection{Infrastructure and Tested DRAM Chips}
\label{sec:infrastructure}

To collect real-chip measurements to fit and validate our model, we use an FPGA-based DRAM testing infrastructure developed from DRAM Bender~\cite{olgun2023dram, safari-drambender} that gives us fine-grained control over DRAM commands, timings, data, and temperature. Table~\ref{tab:modules} lists the 24 commercial off-the-shelf DDR4 modules we test with DRAM chips from five different manufacturers and spans a wide range of DRAM die densities and revisions. We also test an HBM2 stack on a Virtex XCU55 UltraScale+ FPGA from an AMD Xilinx Alveo U55C FPGA board. We reverse engineer the 1) internal DRAM row mapping, and 2) true- and anti-cell layout following the same methodologies in prior works~\cite{kim2020revisiting, orosa2021deeper, yaglikci2022understanding, luo2023rowpress, olgun2024read, liu2013experimental}.

\begin{table}[t]
\caption{DRAM chips tested.}
\label{tab:modules}
\centering
\scriptsize
\setlength{\tabcolsep}{5pt}
\renewcommand{\arraystretch}{0.92}
\resizebox{\columnwidth}{!}{%
\begin{tabular}{@{}c|c|cc|c|c|c@{}}
\toprule
{\textbf{Mfr.}} &
{\textbf{ID}} &
{\textbf{Die Density}} &
{\textbf{Die Revision}} &
{\textbf{DQ}} &
{\textbf{Num. Chips}} &
\begin{tabular}[c]{@{}c@{}}\textbf{Date Code}\\\textbf{(YYWW)}\end{tabular}
 \\ \midrule
\multirow{7}{*}{Samsung (S)} & S00 & 4Gb & E & $\times$8 & 8 & -- \\
 & S01 & 4Gb & F & $\times$8 & 8 & -- \\
 & S02 & 8Gb & B & $\times$8 & 8 & 1617 \\
 & S03 & 8Gb & D & $\times$8 & 8 & 2110 \\
 & S04 & 16Gb & A & $\times$8 & 8 & 2302 \\
 & S05 & 16Gb & B & $\times$8 & 8 & 2315 \\
 & S06 & 16Gb & M & $\times$8 & 8 & 2118 \\ \midrule
\multirow{9}{*}{SK hynix (H)} & H00 & 4Gb & A & $\times$8 & 8 & -- \\
 & H01 & 8Gb & A & $\times$8 & 8 & 1843 \\
 & H02 & 8Gb & C & $\times$8 & 8 & 2120 \\
 & H03 & 8Gb & D & $\times$8 & 8 & 1938 \\
 & H04 & 8Gb & D & $\times$8 & 8 & 1938 \\
 & H05 & 8Gb & D & $\times$8 & 8 & 1938 \\
 & H06 & 8Gb & M & $\times$8 & 8 & 1616 \\
 & H07 & 16Gb & A & $\times$8 & 8 & 2003 \\
 & H08 & 16Gb & C & $\times$8 & 8 & 2136 \\ \midrule
\multirow{4}{*}{Micron (M)} & M00 & 4Gb & B & $\times$8 & 8 & -- \\
 & M01 & 16Gb & B & $\times$8 & 8 & -- \\
 & M02 & 16Gb & E & $\times$16 & 4 & -- \\
 & M03 & 16Gb & F & $\times$8 & 8 & 2412 \\ \midrule
\multirow{2}{*}{Nanya (N)} & N00 & 8Gb & C & $\times$8 & 8 & 2046 \\
 & N01 & 8Gb & C & $\times$8 & 8 & 2141 \\ \midrule
\multirow{2}{*}{CXMT (C)} & C00 & 8Gb & A & $\times$8 & 8 & 2423 \\
 & C01 & 16Gb & A & $\times$8 & 8 & 2533 \\ \midrule
\multicolumn{2}{c|}{HBM2 (U55C)} & 8Gb & -- & -- & -- & -- \\ \bottomrule
\end{tabular}}
\end{table}

\begin{figure*}[t]
\centering
\includegraphics{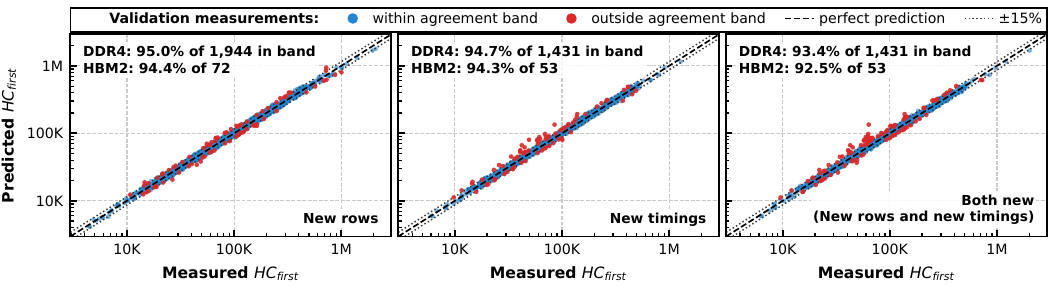}
\caption{Agreement between \hcfirst measured from real DRAM chips and predicted by our model, for all validation measurements.}
\label{fig:heldout-scatter}
\end{figure*}

\subsection{Real-Chip Measurements}
\label{sec:patterns}

We perform two kinds of real-chip measurements: 1) \emph{fitting measurements} that we use to fit the parameters of our model, and 2) \emph{validation measurements} that we use to validate the accuracy of our model's \hcfirst predictions.

\head{Fitting Measurements} We sample 32 \emph{fitting rows} uniformly from a bank of each module (HBM2 stack), excluding rows at the subarray boundary. We measure three types of access patterns: double-sided with alternating aggressor activations, and single-sided with the lower or the upper adjacent row as aggressor. For each access pattern, we also measure with four victim-aggressor data patterns: \vzero, \vone, V1/A1, and V0/A0. For every combination of access and data pattern, we also sweep \tAggON and \tAggOFF. All together, we measure 72 \emph{conditions} (i.e., combinations of access patterns, timings, and data patterns) on the fitting rows of each module, and we fit \model to each module using only these measurements.

\head{Validation Measurements} To validate the \hcfirst prediction accuracy of our model, we first \emph{freeze} the parameters of our model, and then perform three kinds of \emph{new} validation measurements.  1)~\emph{New rows}: the same 72 fitting conditions measured on 32 \emph{validation rows} different from the rows used for fitting. They test whether a model fitted to some rows predicts the other rows of the same module. 2)~\emph{New timings}: 53 conditions that the fitting measurements do not include, measured on the fitting rows. They test whether the model predicts timings it was not fitted to: 40 of them have timings inside the fitted range, 9 have timings beyond it (e.g., \tAggOFF of 340.5\,ns), and 4 have both \tAggON and \tAggOFF above their minimum, which the fitting measurements never vary together. 3)~\emph{Both new}: the same 53 conditions measured on the validation rows. In total, each module (HBM2 stack) thus has 178 validation measurements.

\subsection{Metrics}
\label{sec:metric}

\head{\hcfirst} We measure, fit, and validate \hcfirst, the minimum aggressor row activation count to induce at least one bitflip (counting both aggressor rows for double-sided patterns). We test each activation count three times, and a row's \hcfirst lies between the largest tested count at which no repetition flips a bit and the smallest count at which every repetition does. For each \emph{measurement}, we use the median \hcfirst of the fitting or the validation rows under one condition. We estimate this median and its standard error $s_e$ by maximum likelihood, treating each row's range as an interval-censored sample of a log-normal distribution. If fewer than half of the rows flip within the stimulus window, the measurement gives only the largest tested count instead of a median.

\head{Agreement Criterion} A predicted \hcfirst \emph{agrees} with a measured median \hcfirst if it falls within the median's 95\% confidence interval, slightly widened by a fixed tolerance $\tau$ to account for the natural variation in \hcfirst, i.e.,
\[
  |\ln(\text{predicted}/\text{measured})| \leq 1.96\sqrt{s_e^2 + \tau^2}.
\]
We use $\tau = 0.055$ (5.7\%) for both the fitting and the validation measurements. With the median standard error of our measurements ($s_e = 0.043$), this \emph{agreement band} (i.e., the range of predicted \hcfirst values that agree with a measured median) spans about $\pm$15\% of the measured median.

\section{Model Accuracy Validation Results}
\label{sec:validation}

We evaluate \model in three steps. First, we check how well it reproduces the measurements it is fitted to (Section~\ref{sec:val-fit}). Second, we compare its saved predictions with the validation measurements (Section~\ref{sec:val-heldout}). Third, we analyze where the predictions miss (Section~\ref{sec:val-misses}). Each step includes the HBM2 stack, which we test with the same methodology.

\subsection{Fitting Accuracy}
\label{sec:val-fit}

We first evaluate how accurate our model's predicted \hcfirst is on the fitting measurements. We make two observations. First, for all DDR4 modules, 98.3\% of the fitted model's predicted \hcfirst agrees with the real-chip measurements (i.e., within the 15\% agreement band). On the HBM2 stack, 93.1\% of the fitted model's predicted \hcfirst agrees with the real-chip measurements. Second, it is necessary to fit our model individually to each DRAM module. If we freeze every parameter fitted to a single module \modref{S01}, and refit only the disturbance scale $s$ (Table~\ref{tab:parameters}) to each other module, only 55.5\% of the model's predicted \hcfirst agrees with real-chip measurements.

\observation{When fitted to a DRAM module, our model accurately reproduces the \hcfirst measurements that it is fitted with.}

\subsection{Prediction Accuracy on Validation Data}
\label{sec:val-heldout}

We next evaluate how accurately our fitted model predicts \hcfirst on the validation measurements. Figure~\ref{fig:heldout-scatter} shows the agreement between the measured (x-axis) and model predicted (y-axis) \hcfirst of all validation measurements for new rows, new timings, and both new rows and timings beyond the fitting measurements. We make two observations. First, for all DDR4 modules, 94.4\% of our model's predicted \hcfirst agrees with the real-chip measurements. The median error is 5.4\%. On the HBM2 stack, 93.8\% of the predicted \hcfirst agrees (167 of 178), with a median error of 5.1\%. Second, our model is as accurate on new rows (95.0\% agree) as at new timings (94.7\%) and when both are new (93.4\%); on the HBM2 stack, 68 of 72, 50 of 53, and 49 of 53 predictions agree.

\begin{figure*}[t]
\centering
\includegraphics{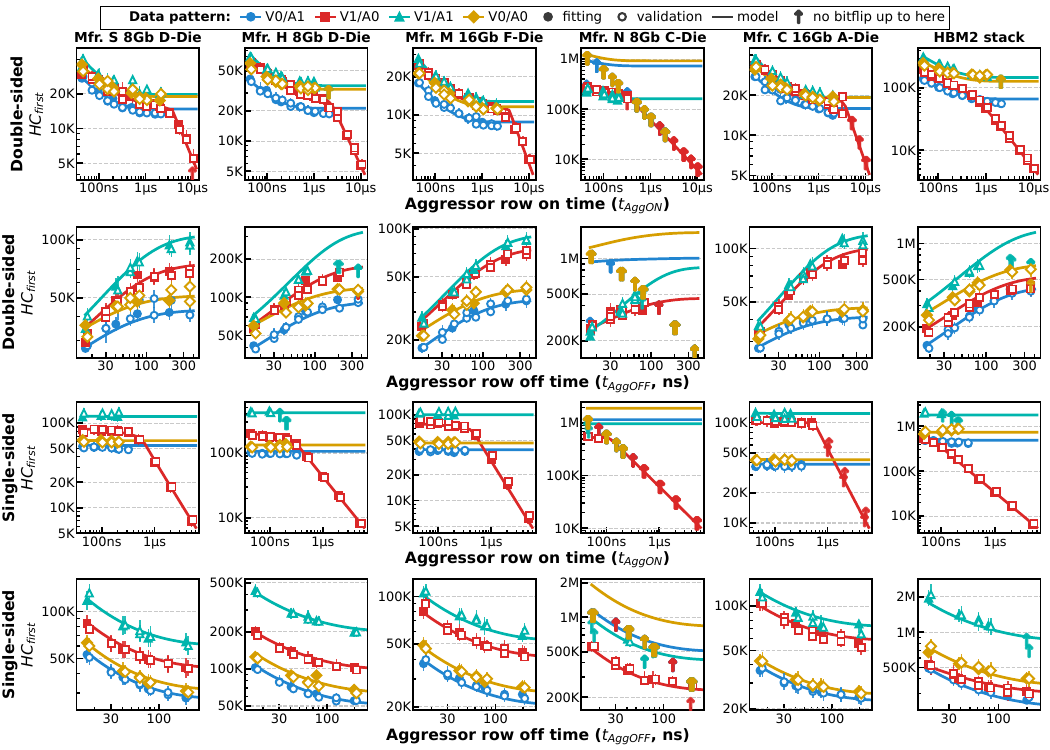}
\caption{Measured and predicted \hcfirst (y-axis) for representative modules and the HBM2 stack at different access patterns (different rows of plots), \tAggON and \tAggOFF (x-axis), and data patterns (different colors in each plot).}
\label{fig:curves}
\end{figure*}

\observation{Our model accurately predicts the \hcfirst measurements that it is \emph{not} fitted with.}

Specifically, we also examine whether our model predicts timings beyond the fitted range. Of the measurements at timings beyond the fitted range (e.g., a \tAggOFF of 340.5\,ns, whereas the fitting measurements reach at most 196.5\,ns), 95.9\% agree, compared to 95.4\% inside the range. On the HBM2 stack, 16 of the 18 such measurements agree (88.9\%), compared to 96.2\% inside the range. The agreement is also similar across access patterns: 94.8\% to 96.7\% for the double- and single-sided \tAggON and \tAggOFF sweeps and for the upper aggressor (HBM2: 90.6\% to 100\%).

\observation{Our model predicts timings beyond the fitted range as accurately as timings inside it.}

As visual examples, Figure~\ref{fig:curves} shows the measured and predicted \hcfirst of one module from each of the five manufacturers and of the HBM2 stack as the access pattern, \tAggON, \tAggOFF, and the data pattern change. Filled markers show the fitting measurements, hollow markers the validation measurements, and lines the fitted model; an upward arrow marks a measurement where no bitflip is observed under the measurement's condition.

\takeaway{Our architecture-level error model of DRAM read disturbance accurately predicts \hcfirst for 1) single- and double-sided access patterns, 2) a wide range of \tAggON and \tAggOFF, and 3) different data patterns.}

\subsection{Analysis of Prediction Errors}
\label{sec:val-misses}

\begin{figure*}[t]
\centering
\includegraphics{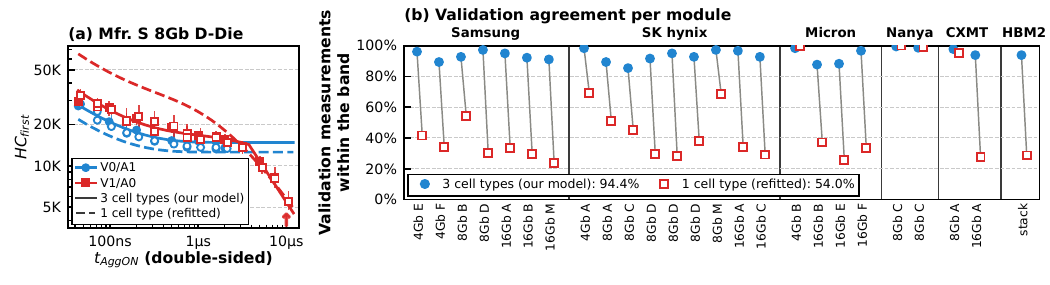}
\caption{(a) Example visualization of \hcfirst vs.\ \tAggON of module \modref{S03} (double-sided), with \model fitted with tail and bulk cell types (solid) and with one cell type (dashed), fitting (filled) and validation (hollow) measurements; (b) share of agreeing validation measurements per module and for the HBM2 stack, for both fits.}
\label{fig:ablation}
\end{figure*}

We find that the largest prediction errors come from conditions with both a long \tAggON and a long \tAggOFF: 19 of the 20 largest errors are from the 216 validation measurements of the four such conditions (Section~\ref{sec:patterns}). Only 75.9\% of these measurements agree, compared to 94.4\% of all validation measurements, and \model predicts on average 12.2\% more activations than the chips need. For example, at a \tAggON of 499.5\,ns and a \tAggOFF of 196.5\,ns, 41 of 108 measurements miss in this direction. The HBM2 stack shows the same miss: at 501.7\,ns on and 198.3\,ns off, \model predicts 39\%--40\% more activations than the chip needs (\vzero). We provide two hypotheses for these errors: 1) the fitting measurements vary \tAggON and \tAggOFF one at a time, so they constrain each effect separately, but not how the two combine; and 2) existing device-level mechanisms do \emph{not} fully explain when both \tAggON and \tAggOFF are long.

\section{Cell Type Ablation Study}
\label{sec:insights}
\label{sec:insights-ablation}

One of the key modeling methodologies in our architecture-level DRAM read disturbance error model is to represent a (victim) DRAM row with three distinct cell types (Section~\ref{sec:model-cell-types}). To test whether this methodology is key to the accuracy of our model, we modify our model to include \emph{only one} victim cell type (i.e., all device-level mechanisms act on one shared state) and then refit this one-cell-type model to the DDR4 modules and the HBM2 stack using the same parameter fitting methodology. Figure~\ref{fig:ablation} shows (a) the double-sided \tAggON sweep of one example module with both the three-cell-type and the one-cell-type models, and (b) the share of agreeing validation measurements per module and for the HBM2 stack.

\observation{Without distinct cell types, the model neither reproduces the fitting measurements nor predicts the validation measurements.}

We observe that the \hcfirst values predicted by the one-cell-type model agree with only 52.3\% of the fitting measurements and 54.0\% of the validation measurements, compared to 98.3\% and 94.4\% with the three-cell-type model. On the HBM2 stack, the one-cell-type model agrees with only 23.6\% of the fitting measurements and 28.7\% of the validation measurements, compared to 93.1\% and 93.8\% with three cell types. On the DDR4 modules, the median validation error of the one-cell-type model is 25.8\%, compared to 5.4\% for the three-cell-type model.

\observation{One cell type cannot accurately produce both bitflip directions.}

In one victim DRAM cell, the \NWL-side traps drive the state toward 1-to-0 bitflips and the \PWL-side traps toward 0-to-1 bitflips, so the two directions compete for the same state. The parameter fitting settles on a compromise between the two competitive mechanisms (Figure~\ref{fig:ablation}a): at a short \tAggON, it predicts the 0-to-1 bitflips of \vzero too early and the 1-to-0 bitflips of \vone too late, while at a long \tAggON, where the leakage of the bulk cell alone sets \hcfirst, its \vone predictions come close again.

\takeaway{Distinct cell types are essential to accurately model DRAM read disturbance.}

\section{Use Case I: Reproducing RowPress in gem5}
\label{sec:use-case}

Existing architecture-level DRAM read disturbance error models integrated into full system simulators like gem5~\cite{binkert2011gem5} do not model RowPress because they do not take \tAggON into account (Section~\ref{sec:related}). To this end, we execute the user-level program that induces RowPress bitflips in prior work~\cite{luo2023rowpress} in gem5 with Ramulator~2~\cite{luo2023ramulator2, luo2026ramulator21, ramulator2github} in full-system mode and feed the DRAM command trace into our model to demonstrate that our model can capture the key characteristics of RowPress bitflips in a full system simulation.

\subsection{Methodology}
\label{sec:use-case-methodology}

\head{Simulated System Configuration} We run the program in gem5 v25.1.0.1~\cite{binkert2011gem5} in full-system mode, booting the Linux 6.8.0-52-generic kernel on the Ubuntu 24.04 disk image of gem5 resources (versions 1.0.0 and 4.0.0)~\cite{bruce2021enabling, gem5resources}, with Ramulator~2.1~\cite{luo2023ramulator2, luo2026ramulator21, ramulator2github} as the memory controller and DRAM model. Table~\ref{tab:use-case-system} shows the configuration.

\begin{table}[t]
\centering
\caption{Simulated system configuration.}
\label{tab:use-case-system}
\footnotesize
\setlength{\tabcolsep}{4pt}
\resizebox{\columnwidth}{!}{%
\begin{tabular}{@{}ll@{}}
\toprule
\textbf{Processor} & 1 core, x86, 3\,GHz, out of order \\
\textbf{Caches} & private L1-I/D 32\,KiB, 8-way; private L2 1\,MiB, 16-way \\
\textbf{Memory controller} & FR-FCFS, open row, all-bank refresh \\
 & ($t_{\mathrm{REFI}}$\,=\,7.8\,\textmu s, $t_{\mathrm{RFC}}$\,=\,350\,ns), row:bank:column mapping \\
\textbf{DRAM} & DDR4-2400 (16-16-16)~\cite{jedec2020ddr4}, 1 channel, 1 rank, 8Gb $\times$8 chips, \\
 & 16 banks, 65,536 rows per bank \\
\textbf{Operating system} & Linux 6.8 (Ubuntu 24.04), 512 huge pages of 2\,MiB \\
\bottomrule
\end{tabular}}
\end{table}

\head{Workload} We adapt\footnote{We port the program to the simulated system by 1)~adapting it to the simulated memory controller's address mapping, 2)~replacing AVX stores with SSE2 stores, since gem5 does not implement AVX, and 3)~allocating 2\,MiB instead of 1\,GiB huge pages, since gem5's x86 model does not support 1\,GiB pages.} the proof-of-concept program that~\cite{luo2023rowpress} uses to demonstrate RowPress bitflips in a real system. The program contains a ``hammering loop'' where in each iteration, the program 1) activates two aggressor rows in a double-sided access pattern, and 2) reads $N$ cache lines from the aggressor row per activation (i.e., $N = 1$: RowHammer; $N > 1$: RowPress; we sweep $N$ from 1 to 64) to increase \tAggON. The program then activates 16 other rows of the same bank to 1) close the aggressor rows, and 2) bypass the TRR in-DRAM RowHammer mitigation~\cite{frigo2020trrespass, hassan2021utrr}, and synchronizes itself with (i.e., waits for) the next refresh command using memory access latency timing side channels. We run the hammering loop for two full refresh windows of the victim row.

\head{Model Configuration} For the victim row, we turn the commands to its bank into open and close events of its two adjacent wordlines (Section~\ref{sec:model}) and evaluate our model with the fitted parameters of each of the 24 DDR4 modules (Section~\ref{sec:model-fitting}). We evaluate both \vzero and \vone data patterns and both arrangements of the two adjacent rows (i.e., \NWL below and \PWL above the victim, and vice versa). We assume that each all-bank refresh command refreshes the next 8 rows in a fixed order and that refreshing the victim row resets the state of all its three cell types.

\head{Metric} For each module and $N$, we report the victim row's \emph{normalized disturbance}, i.e., the accumulated disturbance of its most disturbed cell type when the row is refreshed normalized by the failure threshold $\theta$ (Section~\ref{sec:model-hcfirst}). We take the largest normalized disturbance value over both data patterns and both aggressor arrangements. A normalized disturbance larger than or equal to 1 means that our model predicts at least one bitflip in the victim row before it is refreshed.

\subsection{Key Results}
\label{sec:use-case-results}

\begin{figure}[t]
\centering
\includegraphics{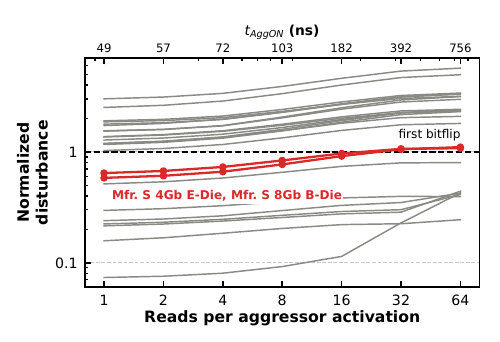}
\caption{Normalized disturbance of the victim row (y-axis) vs.\ the reads per aggressor activation (i.e., \tAggON, x-axis) of the RowPress proof-of-concept program in gem5 for all modules we test. A normalized disturbance equal to or larger than 1.0 (dashed horizontal line) means our model predicts a bitflip before the victim row is refreshed.}
\label{fig:use-case}
\end{figure}

Figure~\ref{fig:use-case} shows the normalized disturbance of the victim row (y-axis) for each module as $N$ (the number of cache line reads per aggressor row activation, x-axis) increases. We make two observations. First, all modules follow the expected trend: the normalized disturbance increases as $N$ increases, i.e., as the program keeps the aggressor rows open for longer at the same number of activations. From one to 32 reads, the normalized disturbance increases by 1.33$\times$ (module \modref{N00}) to 3.11$\times$ (module \modref{M00}). Second, for module \modref{S00} and module \modref{S02}, our model \emph{reproduces} the result of the RowPress paper~\cite{luo2023rowpress} that the program induces bitflips leveraging RowPress (i.e., with higher $N$) while RowHammer cannot.

\takeaway{Our model reproduces the key RowPress bitflip characteristics using DRAM command traces from a real program used to induce RowPress bitflips in full-system simulation.}

\section{Use Case II: More Accurately Accounting for RowPress in Mitigations}
\label{sec:use-case-mitigation}

A RowPress-aware read-disturbance mitigation technique needs to accurately configure itself based on how \hcfirst decreases as \tAggON increases to be both effective (i.e., no false negatives) and efficient (i.e., fewer false positives). DSAC~\cite{hong2023dsac} and ImPress~\cite{qureshi2024impress} account for RowPress by converting an activation that keeps the aggressor row open for \tAggON into a number of \emph{equivalent activations} (EACT):
\[
  \mathrm{HC}_{\mathrm{first}}(t_{\mathrm{AggON}}) = \frac{\mathrm{HC}_{\mathrm{first}}(t_{\mathrm{RAS}})}{\mathrm{EACT}(t_{\mathrm{AggON}})}.
\]
Our goal is to evaluate how accurately these models and our model predict \hcfirst as \tAggON increases.

\head{Methodology} We compare three models:

1)~ImPress, which uses
\[
  \mathrm{EACT} = 1 + \alpha\,\frac{t_{\mathrm{AggON}} - t_{\mathrm{RAS}}}{t_{\mathrm{RC}}},
\]
with both their published $\alpha = 0.48$, and $\alpha$ values fitted individually to each of our tested modules.

2)~DSAC, which uses
\[
  \mathrm{EACT} = 1 + \alpha \log_2\!\left(\frac{t_{\mathrm{AggON}}}{t_{\mathrm{RAS}}}\right)
\]
with $\alpha$ fitted individually to each of our tested modules (the DSAC paper~\cite{hong2023dsac} does not publish an $\alpha$ value).

3)~our model, fitted individually to each of our tested modules.

We use $t_{\mathrm{RAS}} = 43.5$\,ns and $t_{\mathrm{RC}} = 60$\,ns (our minimum aggressor on time and activation period), and we score all models on the same validation measurements with \tAggON up to the 7.8\,\textmu s refresh interval, using the same agreement criterion (Section~\ref{sec:metric}).

\begin{figure}[t]
\centering
\includegraphics{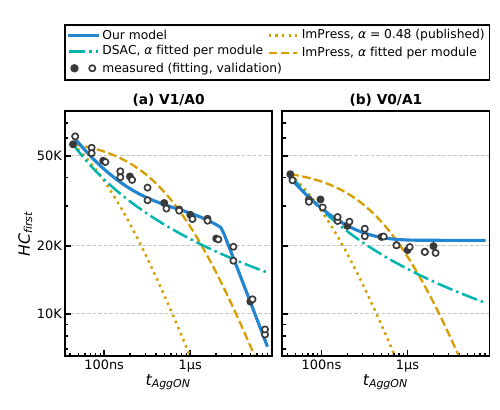}
\caption{Double-sided \hcfirst (y-axis) vs.\ \tAggON (x-axis) of an example module \modref{H04}, measurements (fitting: filled; validation: hollow) and predictions by our model (solid lines), ImPress with the published $\alpha$, and ImPress and DSAC with $\alpha$ fitted to the module; (a)~data pattern V1/A0, (b)~data pattern V0/A1.}
\label{fig:rowpress-models}
\end{figure}

\begin{table}[t]
\caption{Median error of the predicted \hcfirst at each \tAggON (minimum \tAggOFF) over the validation measurements with a median of all 24 DDR4 modules. Positive: too many activations (unsafe); negative: too few (conservative).}
\label{tab:rowpress-models}
\centering
\footnotesize
\setlength{\tabcolsep}{4pt}
\renewcommand{\arraystretch}{0.92}
\begin{tabular*}{\columnwidth}{@{\extracolsep{\fill}}r*{4}{r}@{}}
\toprule
& \multicolumn{4}{c}{\textbf{Median error of the predicted \hcfirst}} \\
\cmidrule(lr){2-5}
\tabhead{\textbf{\tAggON}\\\textbf{(ns)}} & \tabhead{\textbf{Our}\\\textbf{model}} & \tabhead{ImPress\\$\alpha$ = 0.48} & \tabhead{ImPress\\fitted $\alpha$} & \tabhead{DSAC\\fitted $\alpha$} \\
\midrule
43.5 & \textbf{\boldmath +1.33\%} & $-$0.62\% & $-$0.62\% & $-$0.62\% \\
67.5 & \textbf{\boldmath +0.27\%} & $-$10.45\% & +3.42\% & $-$5.30\% \\
97.5 & \textbf{\boldmath $-$1.28\%} & $-$26.27\% & +0.11\% & $-$14.83\% \\
145.5 & \textbf{\boldmath $-$0.97\%} & $-$28.05\% & +15.92\% & $-$4.94\% \\
199.5 & \textbf{\boldmath $-$1.44\%} & $-$46.36\% & +2.91\% & $-$19.62\% \\
301.5 & \textbf{\boldmath +0.11\%} & $-$49.05\% & +19.95\% & $-$2.45\% \\
499.5 & \textbf{\boldmath $-$4.57\%} & $-$62.06\% & +24.10\% & $-$1.08\% \\
703.5 & \textbf{\boldmath $-$2.87\%} & $-$70.55\% & +15.95\% & +1.54\% \\
1003.5 & \textbf{\boldmath +0.68\%} & $-$76.72\% & +9.92\% & +3.55\% \\
1501.5 & \textbf{\boldmath +5.31\%} & $-$83.26\% & $-$5.53\% & +3.54\% \\
2005.5 & \textbf{\boldmath +6.18\%} & $-$86.72\% & $-$16.22\% & +3.47\% \\
3001.5 & \textbf{\boldmath +9.05\%} & $-$89.45\% & $-$31.61\% & $-$0.12\% \\
5005.5 & \textbf{\boldmath +0.29\%} & $-$71.35\% & +113.33\% & +321.92\% \\
7003.5 & \textbf{\boldmath $-$4.68\%} & $-$90.59\% & $-$41.59\% & +64.59\% \\
\midrule
\textbf{Agree} & \textbf{\boldmath 95.0\%} & 33.5\% & 69.3\% & 71.2\% \\
\bottomrule
\end{tabular*}

\end{table}

Table~\ref{tab:rowpress-models} lists the median error of the predicted \hcfirst at each \tAggON over all 24 DDR4 modules. We make two observations.

First, our model is more accurate than both prior models: 95.0\% of its predictions agree with the 2,835 validation measurements of the \tAggON sweeps (median error 5.5\%), compared to 69.3\% for ImPress and 71.2\% for DSAC with $\alpha$ fitted to each module (median errors 11.9\% and 11.5\%), and 33.5\% for ImPress with the published $\alpha$ (median error 47.2\%).

Second, a single equivalent-activation curve cannot accurately follow both the RowHammer (i.e., \tAggON $= t_{\mathrm{RAS}}$) and the RowPress (i.e., \tAggON $> t_{\mathrm{RAS}}$) regimes of \hcfirst. Figure~\ref{fig:rowpress-models} shows the measured and predicted \hcfirst as \tAggON increases for one example module: the measured \hcfirst decreases slowly up to about 2--3\,\textmu s and rapidly beyond, where the leakage of the bulk cells takes over from the tail cells (Figure~\ref{fig:rowpress-models}a).

We also find that DSAC's logarithmic curve follows the slow decrease but then decreases too slowly: at 5--7\,\textmu s, it predicts 64.59\% to 321.92\% more activations than the chips need (Table~\ref{tab:rowpress-models}), so a mitigation configured with it would act too late. ImPress's linear curve deviates in both directions: with a fitted $\alpha$, it predicts up to 24.10\% more activations at 145--700\,ns and up to 31.61\% fewer at 2--3\,\textmu s; with the published $\alpha$, it predicts 76.72\% to 89.45\% fewer activations than measured at 1--3\,\textmu s, so a mitigation configured with it acts far more often than needed.

Both prior models also scale every data pattern by the same curve, whereas the measured \hcfirst of V0/A1 stops decreasing above 1\,\textmu s while that of V1/A0 keeps decreasing (Figure~\ref{fig:rowpress-models}b); our model follows both, because the leakage that dominates at long on times drives only 1-to-0 bitflips (Section~\ref{sec:model-mechanisms}).

\takeaway{Our model predicts how \hcfirst decreases as \tAggON increases much more accurately than the modeling of RowPress in prior mitigation works.}

\section{Related Work}
\label{sec:related-work}

To our knowledge, this is the first work to build an architecture-level error model of DRAM read disturbance that is principled on device-level mechanisms, fitted to real DRAM chips, and validated on a wide range of conditions (i.e., DRAM access patterns, timings, and data patterns).

\head{Error Models of DRAM Read Disturbance} We compare our model in detail with prior architecture-level models~\cite{thomas2023hammulator, goswami2026hammersimsystemleveltoolmodel, france2021implementing, france2022modeling}, the RowPress models of read-disturbance mitigations~\cite{hong2023dsac, qureshi2024impress}, and circuit-level models~\cite{demirel2025circuit, jiang2021quantifying, hu2026quantifying} in Section~\ref{sec:related} (Table~\ref{tab:related-models}). Device-level works explain the mechanisms of read disturbance with TCAD simulations and wafer-level measurements~\cite{ryu2017overcoming, yang2019trap, walker2021ondramrowhammer, Gautam2020MitigatingPassing, zhou2023double, Zhou2024Unveiling, Zhou2024Understanding, luo2026demystifying,bai2026unified, park2014active, park2016experiments, yang2016suppression, gautam2019rowhammering, han2021surround} (Section~\ref{sec:bg-mechanisms}). Our model builds on these device-level mechanisms to form a practical DRAM read disturbance error model usable by computer architects.

\head{Experimental Characterization of DRAM} Many works experimentally characterize real DRAM chips, e.g., their data retention~\cite{liu2013experimental, khan2014efficacy, patel2017reaper}, access latency~\cite{chang2016understanding, lee2017design, kim2018solar}, and reduced-voltage operation~\cite{chang2017understanding}. Many works characterize RowHammer~\cite{kim2014flipping, kim2020revisiting, orosa2021deeper, yaglikci2022understanding, yaglikci2024spatial, olgun2025variable, tugrul2025understanding, wang2026scaledisturb, park2016statistical, park2016experiments, lim2017active, lim2018study, yun2018study, baeg2022estimation, lang2023blaster, he2023whistleblower, orosa2024spyhammer}, RowPress~\cite{luo2023rowpress, luo2024rowpress, luo2024experimental}, and ColumnDisturb~\cite{yuksel2025columndisturb} in real DDR3, DDR4, and HBM2~\cite{olgun2023hbm, olgun2024read} chips. These works only empirically characterize DRAM read disturbance bitflips (e.g., measure \hcfirst) under their tested conditions, and they do \emph{not} propose architecture-level error modeling of DRAM read disturbance principled on device-level mechanisms.

\head{Reverse Engineering of DRAM} Prior works reverse engineer the internals of commodity DRAM chips, e.g., row mappings and internal structure~\cite{kim2014flipping, kim2020revisiting, orosa2021deeper,luo2023rowpress, nam2023xray, nam2024dramscope}, true- and anti-cell layouts~\cite{kim2014flipping, liu2013experimental}, sense-amplifier circuits~\cite{marazzi2024hifidram}, on-die ECC functions~\cite{patel2019understanding, patel2020beer}, and in-DRAM read-disturbance mitigations~\cite{frigo2020trrespass, hassan2021utrr}. However, these works do not model DRAM read disturbance. We follow these methodologies to reverse engineer the DRAM internal structures required in our work (e.g., the DRAM chip internal row mapping and the true- and anti-cell layout of the tested chips).

\head{Mitigating DRAM Read Disturbance} Many works mitigate read-disturbance bitflips by preventively refreshing potential victim rows~\refreshBasedRowHammerDefenseCitations{} or by throttling the activations of aggressor rows~\cite{yaglikci2021blockhammer, canpolat2024breakhammer}. These mitigations are typically configured with a single activation threshold per system. Some RowPress-aware mitigations count an activation that keeps the aggressor row open for long as multiple equivalent activations~\cite{hong2023dsac, qureshi2024impress} or limit the aggressor row on time~\cite{luo2023rowpress}. Our modeling work is complementary to the DRAM read disturbance mitigation mechanisms and can help to configure the mitigation mechanisms more accurately and efficiently (Section~\ref{sec:use-case-mitigation}).

\section{Conclusion}
\label{sec:conclusion}

In this paper, we build a principled architecture-level error model for DRAM read disturbance and validate it on a wide range of real DRAM chips. Our model abstracts a victim row into three cell types whose states respond to the opening and closing of the two adjacent wordlines through the mechanisms identified by device-level works. The parameters of our model can be fitted to individual DRAM modules.

Our evaluation using real-chip \hcfirst measurements from 24 commercial off-the-shelf DDR4 modules with DRAM chips from five manufacturers and one HBM2 stack shows that our model accurately reproduces the fitting measurements and predicts \hcfirst on new victim rows and at new timings, including timings beyond the fitted range, and it is equally accurate on the tested HBM2 stack. Our ablation study shows that distinguishing victim cell types is essential to the accuracy of the model.

We demonstrate two use cases of our model. First, our model reproduces the key characteristics of RowPress bitflips in full-system simulation of a real program that prior works use to induce RowPress bitflips. Second, our model predicts how \hcfirst decreases as \tAggON increases much more accurately than the RowPress models of prior mitigations.

We hope that our model 1)~enables a more accurate evaluation of a system's vulnerability to DRAM read disturbance, 2)~helps design and configure safer and more efficient DRAM read disturbance mitigations, and 3)~serves as a foundation for future works to fundamentally understand and model DRAM read disturbance, at both the device and the architecture level.

\section*{Generative AI and LLM Usage Acknowledgment}
\label{sec:ai-usage}

In this project, we used Claude Opus~5.5 and Claude Mythos~5.1 and~5.2 for 1)~generic coding assistance, 2)~streamlining and automating experiment workflows, 3)~drafting figures and plots, 4)~language editing, and 5)~critiquing and optimizing the analytical modeling of the device-level error mechanisms of DRAM read disturbance. The authors reviewed all generated content and take full responsibility for the paper.

\section*{Acknowledgments}
We thank Prof. Longda Zhou for our earlier collaboration on understanding the device-level physical mechanisms of DRAM
read disturbance~\cite{luo2026demystifying}. We also thank M{\"u}kerrem T{\"u}fek{\c{c}}io{\u{g}}lu and Prof. Yanjing Li
for our early collaborative efforts exploring architecture-level error modeling of DRAM read disturbance. We thank
members of the SAFARI Research Group for their constructive feedback and for providing a stimulating, intellectual, and
scientific environment. We acknowledge the generous gift funding provided by our industrial partners (especially Google,
Huawei, Intel, Microsoft), which has been instrumental in enabling the research we have been conducting on read
disturbance in DRAM.

\bibliographystyle{IEEEtran}
\bibliography{combined_refs}

\begin{thebibliography}{100}
\providecommand{\url}[1]{#1}
\csname url@samestyle\endcsname
\providecommand{\newblock}{\relax}
\providecommand{\bibinfo}[2]{#2}
\providecommand{\BIBentrySTDinterwordspacing}{\spaceskip=0pt\relax}
\providecommand{\BIBentryALTinterwordstretchfactor}{4}
\providecommand{\BIBentryALTinterwordspacing}{\spaceskip=\fontdimen2\font plus
\BIBentryALTinterwordstretchfactor\fontdimen3\font minus
  \fontdimen4\font\relax}
\providecommand{\BIBforeignlanguage}[2]{{%
\expandafter\ifx\csname l@#1\endcsname\relax
\typeout{** WARNING: IEEEtran.bst: No hyphenation pattern has been}%
\typeout{** loaded for the language `#1'. Using the pattern for}%
\typeout{** the default language instead.}%
\else
\language=\csname l@#1\endcsname
\fi
#2}}
\providecommand{\BIBdecl}{\relax}
\BIBdecl

\bibitem{kim2014flipping}
Y.~{Kim}, R.~{Daly}, J.~{Kim}, C.~{Fallin}, J.~H. {Lee}, D.~{Lee},
  C.~{Wilkerson}, K.~{Lai}, and O.~{Mutlu}, ``{Flipping Bits in Memory Without
  Accessing Them: An Experimental Study of DRAM Disturbance Errors},'' in
  \emph{ISCA}, 2014.

\bibitem{kim2020revisiting}
J.~S. Kim, M.~Patel, A.~G. Ya\u{g}l{\i}k\c{c}{\i}, H.~Hassan, R.~Azizi,
  L.~Orosa, and O.~Mutlu, ``{Revisiting RowHammer: An Experimental Analysis of
  Modern DRAM Devices and Mitigation Techniques},'' in \emph{ISCA}, 2020.

\bibitem{orosa2021deeper}
L.~Orosa, A.~G. Ya{\u{g}}l{\i}k{\c{c}}{\i}, H.~Luo, A.~Olgun, J.~Park,
  H.~Hassan, M.~Patel, J.~S. Kim, and O.~Mutlu, ``{A Deeper Look into
  RowHammer's Sensitivities: Experimental Analysis of Real DRAM Chips and
  Implications on Future Attacks and Defenses},'' in \emph{MICRO}, 2021.

\bibitem{cojocar2020rowhammer}
L.~Cojocar, J.~Kim, M.~Patel, L.~Tsai, S.~Saroiu, A.~Wolman, and O.~Mutlu,
  ``{Are We Susceptible to Rowhammer? An End-to-End Methodology for Cloud
  Providers},'' in \emph{S\&P}, 2020.

\bibitem{mutlu2017rowhammer}
O.~Mutlu, ``{The RowHammer Problem and Other Issues We May Face as Memory
  Becomes Denser},'' in \emph{DATE}, 2017.

\bibitem{mutlu2019rowhammer}
O.~Mutlu and J.~S. Kim, ``{RowHammer: A Retrospective},'' \emph{IEEE TCAD},
  2020.

\bibitem{mutlu2023fundamentally}
O.~Mutlu, A.~Olgun, and A.~G. Yağlıkçı, ``{Fundamentally Understanding and
  Solving RowHammer},'' in \emph{ASP-DAC}, 2023.

\bibitem{mutlu2023retrospective}
\BIBentryALTinterwordspacing
O.~Mutlu, ``{Retrospective: Flipping Bits in Memory without Accessing Them: An
  Experimental Study of DRAM Disturbance Errors},'' arXiv:2306.16093, 2023.
  [Online]. Available: \url{https://arxiv.org/abs/2306.16093}
\BIBentrySTDinterwordspacing

\bibitem{luo2023rowpress}
H.~Luo, A.~Olgun, A.~G. Ya{\u{g}}l{\i}k{\c{c}}{\i}, Y.~C. Tu{\u{g}}rul,
  S.~Rhyner, M.~B. Cavlak, J.~Lindegger, M.~Sadrosadati, and O.~Mutlu,
  ``{RowPress: Amplifying Read Disturbance in Modern DRAM Chips},'' in
  \emph{ISCA}, 2023.

\bibitem{luo2024rowpress}
------, ``{RowPress Vulnerability in Modern DRAM Chips},'' \emph{IEEE Micro},
  vol.~44, no.~4, pp. 60--69, 2024.

\bibitem{luo2024experimental}
H.~Luo, I.~E. Y{\"u}ksel, A.~Olgun, A.~G. Ya{\u{g}}l{\i}k{\c{c}}{\i},
  M.~Sadrosadati, and O.~Mutlu, ``{An Experimental Characterization of Combined
  RowHammer and RowPress Read Disturbance in Modern DRAM Chips},'' in \emph{DSN
  Disrupt}, 2024.

\bibitem{frigo2020trrespass}
P.~Frigo, E.~Vannacci, H.~Hassan, V.~van~der Veen, O.~Mutlu, C.~Giuffrida,
  H.~Bos, and K.~Razavi, ``{TRRespass: Exploiting the Many Sides of Target Row
  Refresh},'' in \emph{{S\&P}}, 2020.

\bibitem{hassan2021utrr}
H.~Hassan, Y.~C. Tugrul, J.~S. Kim, V.~van~der Veen, K.~Razavi, and O.~Mutlu,
  ``{Uncovering in-DRAM RowHammer Protection Mechanisms: A New Methodology,
  Custom RowHammer Patterns, and Implications},'' in \emph{MICRO}, 2021.

\bibitem{kamadan2025ecc}
N.~Kamadan, W.~Wang, S.~{van Schaik}, C.~Garman, D.~Genkin, and Y.~Yarom,
  ``{ECC.fail: Mounting RowHammer Attacks on DDR4 Servers with ECC Memory},''
  in \emph{USENIX Security}, 2025.

\bibitem{nazaraliyev2025not}
R.~Nazaraliyev, Y.~Zhang, S.~B. Dutta, A.~Marquez, K.~Barker, and
  N.~Abu-Ghazaleh, ``{Not So Refreshing: Attacking GPUs Using RFM Rowhammer
  Mitigation},'' in \emph{USENIX Security}, 2025.

\bibitem{orosa2024spyhammer}
L.~Orosa, U.~Rührmair, A.~G. Yağlıkçı, H.~Luo, A.~Olgun, P.~Jattke,
  M.~Patel, J.~S. Kim, K.~Razavi, and O.~Mutlu, ``{SpyHammer: Understanding and
  Exploiting RowHammer Under Fine-Grained Temperature Variations},'' \emph{IEEE
  Access}, 2024.

\bibitem{woo2025mitigations}
J.~Woo, J.~Qu, G.~Saileshwar, and P.~J. Nair, ``{When Mitigations Backfire:
  Timing Channel Attacks and Defense for PRAC-Based RowHammer Mitigations},''
  in \emph{ISCA}, 2025.

\bibitem{aga2017good}
M.~T. Aga, Z.~B. Aweke, and T.~Austin, ``{When Good Protections Go Bad:
  Exploiting Anti-DoS Measures to Accelerate Rowhammer Attacks},'' in
  \emph{HOST}, 2017.

\bibitem{agarwal2018rowhammer}
S.~Agarwal, H.~Dixit, D.~Datta, M.~Tran, D.~Houssameddine, D.~Shum, and
  F.~Benistant, ``{Rowhammer for Spin Torque based Memory: Problem or not?}''
  in \emph{INTERMAG}, 2018.

\bibitem{aydin2022cyber}
H.~Aydin and A.~Sertba{\c{s}}, ``{Cyber Security in Industrial Control Systems
  (ICS): A Survey of RowHammer Vulnerability},'' \emph{Applied Computer
  Science}, 2022.

\bibitem{barenghi2018software}
A.~Barenghi, L.~Breveglieri, N.~Izzo, and G.~Pelosi, ``{Software-Only Reverse
  Engineering of Physical DRAM Mappings for Rowhammer Attacks},'' in
  \emph{IVSW}, 2018.

\bibitem{bhattacharya2016curious}
S.~Bhattacharya and D.~Mukhopadhyay, ``{Curious Case of Rowhammer: Flipping
  Secret Exponent Bits Using Timing Analysis},'' in \emph{CHES}, 2016.

\bibitem{bhattacharya2018advanced}
------, ``{Advanced Fault Attacks in Software: Exploiting the Rowhammer Bug},''
  in \emph{Fault Tolerant Architectures for Cryptography and Hardware
  Security}, 2018.

\bibitem{bolcskei2025rubicon}
M.~B{\"o}lcskei, P.~Jattke, J.~Wikner, and K.~Razavi, ``{Rubicon: Precise
  Microarchitectural Attacks with Page-Granular Massaging},'' in
  \emph{EuroS\&P}, 2025.

\bibitem{bosman2016dedup}
E.~Bosman, K.~Razavi, H.~Bos, and C.~Giuffrida, ``{Dedup Est Machina: Memory
  Deduplication as An Advanced Exploitation Vector},'' in \emph{S\&P}, 2016.

\bibitem{bostanci2025understanding}
F.~N. Bostanc{\i}, O.~Canpolat, A.~Olgun, {\.I}.~E. Y{\"u}ksel,
  K.~Kanellopoulos, M.~Sadrosadati, A.~G. Ya{\u{g}}l{\i}k{\c{c}}{\i}, and
  O.~Mutlu, ``{Understanding and Mitigating Covert Channel and Side Channel
  Vulnerabilities Introduced by RowHammer Defenses},'' in \emph{MICRO}, 2025.

\bibitem{brasser2017can}
F.~Brasser, L.~Davi, D.~Gens, C.~Liebchen, and A.-R. Sadeghi, ``{Can't Touch
  This: Software-Only Mitigation Against Rowhammer Attacks Targeting Kernel
  Memory},'' in \emph{USENIX Security}, 2017.

\bibitem{burleson2016invited}
W.~Burleson, O.~Mutlu, and M.~Tiwari, ``{Invited: Who is the Major Threat to
  Tomorrow's Security? You, the Hardware Designer},'' in \emph{DAC}, 2016.

\bibitem{cai2022feasibility}
K.~Cai, Z.~Zhang, and F.~Yao, ``{On the Feasibility of Training-time Trojan
  Attacks through Hardware-based Faults in Memory},'' in \emph{HOST}, 2022.

\bibitem{carre2018openssl}
S.~Carre, M.~Desjardins, A.~Facon, and S.~Guilley, ``{OpenSSL Bellcore's
  Protection Helps Fault Attack},'' in \emph{DSD}, 2018.

\bibitem{chen2025rhammer}
W.~Chen, S.~Tang, Y.~Tang, X.~Luo, Y.~Zhang, and W.~Qiang, ``{$\rho$Hammer:
  Reviving RowHammer Attacks on New Architectures via Prefetching},'' in
  \emph{MICRO}, 2025.

\bibitem{cohen2022hammerscope}
Y.~Cohen, K.~S. Tharayil, A.~Haenel, D.~Genkin, A.~D. Keromytis, Y.~Oren, and
  Y.~Yarom, ``{HammerScope: Observing DRAM Power Consumption Using
  Rowhammer},'' in \emph{CCS}, 2022.

\bibitem{cojocar2019eccploit}
L.~Cojocar, K.~Razavi, C.~Giuffrida, and H.~Bos, ``{Exploiting Correcting
  Codes: On the Effectiveness of ECC Memory Against Rowhammer Attacks},'' in
  \emph{S\&P}, 2019.

\bibitem{deridder2021smash}
F.~de~Ridder, P.~Frigo, E.~Vannacci, H.~Bos, C.~Giuffrida, and K.~Razavi,
  ``{SMASH}: {Synchronized} {Many-Sided} {Rowhammer} {Attacks} from
  {JavaScript},'' in \emph{{USENIX Security}}, 2021.

\bibitem{deridder2025posthammer}
F.~de~Ridder, P.~Jattke, and K.~Razavi, ``{Posthammer: Pervasive Browser-based
  Rowhammer Attacks with Postponed Refresh Commands},'' in \emph{USENIX
  Security}, 2025.

\bibitem{fahr2022effects}
\BIBentryALTinterwordspacing
M.~J. Fahr, ``{The Effects of Side-Channel Attacks on Post-Quantum
  Cryptography: Influencing FrodoKEM Key Generation Using the Rowhammer
  Exploit},'' Master's thesis, University of Arkansas, 2022. [Online].
  Available: \url{https://scholarworks.uark.edu/etd/4598/}
\BIBentrySTDinterwordspacing

\bibitem{fahr2022frodo}
M.~Fahr, Jr., H.~Kippen, A.~Kwong, T.~Dang, J.~Lichtinger, D.~Dachman-Soled,
  D.~Genkin, A.~Nelson, R.~A. Perlner, A.~Yerukhimovich, and D.~Apon, ``{When
  Frodo Flips: End-to-End Key Recovery on FrodoKEM via Rowhammer},'' in
  \emph{CCS}, 2022.

\bibitem{fournaris2017exploiting}
A.~P. Fournaris, L.~Pocero~Fraile, and O.~Koufopavlou, ``{Exploiting Hardware
  Vulnerabilities to Attack Embedded System Devices: A Survey of Potent
  Microarchitectural Attacks},'' \emph{Electronics}, 2017.

\bibitem{frigo2018grand}
P.~Frigo, C.~Giuffrida, H.~Bos, and K.~Razavi, ``{Grand Pwning Unit:
  Accelerating Microarchitectural Attacks with the GPU},'' in \emph{S\&P},
  2018.

\bibitem{genssler2022reliability}
P.~R. Genssler, V.~M. van Santen, J.~Henkel, and H.~Amrouch, ``{On the
  Reliability of FeFET On-Chip Memory},'' \emph{TC}, 2022.

\bibitem{google-project-zero}
M.~Seaborn and T.~Dullien, ``{Exploiting the DRAM Rowhammer Bug to Gain Kernel
  Privileges},''
  \url{https://projectzero.google/2015/03/exploiting-dram-rowhammer-bug-to-gain.html},
  Mar. 2015.

\bibitem{gruss2016rowhammer}
\BIBentryALTinterwordspacing
D.~Gruss, C.~Maurice, and S.~Mangard, ``{Rowhammer.js: A Remote
  Software-Induced Fault Attack in JavaScript},'' in \emph{DIMVA}, 2016.
  [Online]. Available: \url{https://arxiv.org/abs/1507.06955}
\BIBentrySTDinterwordspacing

\bibitem{gruss2018another}
D.~Gruss, M.~Lipp, M.~Schwarz, D.~Genkin, J.~Juffinger, S.~O'Connell,
  W.~Schoechl, and Y.~Yarom, ``{Another Flip in the Wall of Rowhammer
  Defenses},'' in \emph{S\&P}, 2018.

\bibitem{hong2019terminal}
S.~Hong, P.~Frigo, Y.~Kaya, C.~Giuffrida, and T.~Dumitra\c{s}, ``{Terminal
  Brain Damage: Exposing the Graceless Degradation in Deep Neural Networks
  Under Hardware Fault Attacks},'' in \emph{USENIX Security}, 2019.

\bibitem{hu2026gddrhammer}
Y.~Hu, N.~Brown, Y.~Chen, J.~Bakita, T.~Chen, D.~Genkin, and A.~Kwong,
  ``{GDDRHammer: Greatly Disturbing DRAM Rows --- Cross-Component Rowhammer
  Attacks from Modern GPUs},'' in \emph{S\&P}, 2026.

\bibitem{islam2022signature}
S.~Islam, K.~Mus, R.~Singh, P.~Schaumont, and B.~Sunar, ``{Signature Correction
  Attack on Dilithium Signature Scheme},'' in \emph{Euro S\&P}, 2022.

\bibitem{jang2017sgx}
Y.~Jang, J.~Lee, S.~Lee, and T.~Kim, ``{SGX-Bomb: Locking Down the Processor
  via Rowhammer Attack},'' in \emph{SysTEX}, 2017.

\bibitem{jattke2022blacksmith}
\BIBentryALTinterwordspacing
P.~Jattke, V.~van~der Veen, P.~Frigo, S.~Gunter, and K.~Razavi, ``{Blacksmith}:
  {Scalable} {Rowhammering} in the {Frequency} {Domain},'' in \emph{{S\&P}},
  May 2022. [Online]. Available:
  \url{https://comsec.ethz.ch/research/dram/blacksmith/}
\BIBentrySTDinterwordspacing

\bibitem{jattke2024zenhammer}
\BIBentryALTinterwordspacing
P.~Jattke, M.~Wipfli, F.~Solt, M.~Marazzi, M.~B{\"o}lcskei, and K.~Razavi,
  ``{ZenHammer: Rowhammer Attacks on AMD Zen-based Platforms},'' in
  \emph{USENIX Security}, 2024, pp. 1615--1633. [Online]. Available:
  \url{https://www.usenix.org/conference/usenixsecurity24/presentation/jattke}
\BIBentrySTDinterwordspacing

\bibitem{jattke2025mcsee}
P.~Jattke, M.~Marazzi, F.~Solt, M.~Wipfli, S.~Gloor, and K.~Razavi, ``{McSee:
  Evaluating Advanced Rowhammer Attacks and Defenses via Automated DRAM Traffic
  Analysis},'' in \emph{USENIX Security}, 2025.

\bibitem{ji2019pinpoint}
S.~Ji, Y.~Ko, S.~Oh, and J.~Kim, ``{Pinpoint Rowhammer: Suppressing Unwanted
  Bit Flips on Rowhammer Attacks},'' in \emph{ASIACCS}, 2019.

\bibitem{kang2024sledgehammer}
I.~Kang, W.~Wang, J.~Kim, S.~van Schaik, Y.~Tobah, D.~Genkin, A.~Kwong, and
  Y.~Yarom, ``{SledgeHammer: Amplifying Rowhammer via Bank-level
  Parallelism},'' in \emph{USENIX Security}, 2024.

\bibitem{kaur2022work}
A.~Kaur, P.~Srivastav, and B.~Ghoshal, ``{Work-in-Progress: DRAM-MaUT: DRAM
  Address Mapping Unveiling Tool for ARM Devices},'' in \emph{CASES}, 2022.

\bibitem{khan2018analysis}
M.~N.~I. Khan and S.~Ghosh, ``{Analysis of Row Hammer Attack on STTRAM},'' in
  \emph{ICCD}, 2018, pp. 75--82.

\bibitem{kogler2022half}
A.~Kogler, J.~Juffinger, S.~Qazi, Y.~Kim, M.~Lipp, N.~Boichat, E.~Shiu,
  M.~Nissler, and D.~Gruss, ``{Half-Double: Hammering From the Next Row
  Over},'' in \emph{USENIX Security}, 2022.

\bibitem{kwong2020rambleed}
A.~Kwong, D.~Genkin, D.~Gruss, and Y.~Yarom, ``{RAMBleed: Reading Bits in
  Memory Without Accessing Them},'' in \emph{S\&P}, 2020.

\bibitem{lefforge2023reverse}
S.~Lefforge, ``{Reverse Engineering Post-Quantum Cryptography Schemes to Find
  Rowhammer Exploits},'' Undergraduate Honors Thesis, University of Arkansas,
  2023.

\bibitem{li2014write}
H.~Li, H.-Y. Chen, Z.~Chen, B.~Chen, R.~Liu, G.~Qiu, P.~Huang, F.~Zhang,
  Z.~Jiang, B.~Gao, L.~Liu, X.~Liu, S.~Yu, H.-S.~P. Wong, and J.~Kang, ``{Write
  Disturb Analyses on Half-Selected Cells of Cross-Point RRAM Arrays},'' in
  \emph{IRPS}, 2014.

\bibitem{li2022cyberradar}
D.~Li, D.~Liu, Y.~Ren, Z.~Wang, Y.~Sun, Z.~Guan, Q.~Wu, and J.~Liu,
  ``{FPHammer: A Device Identification Framework based on DRAM
  Fingerprinting},'' in \emph{TrustCom}, 2023.

\bibitem{li2025oneflip}
X.~Li, Y.~Meng, J.~Chen, L.~Luo, and Q.~Zeng, ``{Rowhammer-Based Trojan
  Injection: One Bit Flip Is Sufficient for Backdooring DNNs},'' in
  \emph{USENIX Security}, 2025.

\bibitem{lim2017active}
C.~Lim, K.~Park, and S.~Baeg, ``{Active Precharge Hammering to Monitor
  Displacement Damage Using High-Energy Protons in 3x-nm SDRAM},'' \emph{TNS},
  2017.

\bibitem{lin2025gpuhammer}
C.~S. Lin, J.~Qu, and G.~Saileshwar, ``{GPUHammer: Rowhammer Attacks on GPU
  Memories are Practical},'' in \emph{USENIX Security}, 2025.

\bibitem{lin2026gpubreach}
C.~S. Lin, Y.~Yan, G.~Ding, J.~Qu, J.~Zhu, D.~Lie, and G.~Saileshwar,
  ``{GPUBreach: Privilege Escalation Attacks on GPUs using Rowhammer},'' in
  \emph{S\&P}, 2026.

\bibitem{lipp2018nethammer}
\BIBentryALTinterwordspacing
M.~Lipp, M.~Schwarz, L.~Raab, L.~Lamster, M.~T. Aga, C.~Maurice, and D.~Gruss,
  ``{Nethammer: Inducing Rowhammer Faults through Network Requests},'' in
  \emph{EuroS\&PW}, 2020. [Online]. Available:
  \url{https://conferences.computer.org/eurosp/pdfs/EuroSPW2020-7k9FlVRX4z43j4uE2SeXU0/859700a709/859700a709.pdf}
\BIBentrySTDinterwordspacing

\bibitem{liu2022generating}
L.~Liu, Y.~Guo, Y.~Cheng, Y.~Zhang, and J.~Yang, ``{Generating Robust DNN with
  Resistance to Bit-Flip based Adversarial Weight Attack},'' \emph{IEEE TC},
  2023.

\bibitem{meyer2026phoenix}
D.~Meyer, P.~Jattke, M.~Marazzi, S.~Qazi, D.~Moghimi, and K.~Razavi,
  ``{Phoenix: Rowhammer Attacks on DDR5 with Self-Correcting
  Synchronization},'' in \emph{S\&P}, 2026.

\bibitem{mus2022jolt}
K.~Mus, Y.~Dor{\"o}z, M.~C. Tol, K.~Rahman, and B.~Sunar, ``{Jolt: Recovering
  TLS Signing Keys via Rowhammer Faults},'' in \emph{S\&P}, 2023.

\bibitem{ni2018write}
K.~Ni, X.~Li, J.~A. Smith, M.~Jerry, and S.~Datta, ``{Write Disturb in
  Ferroelectric FETs and Its Implication for 1T-FeFET AND Memory Arrays},''
  \emph{IEEE EDL}, 2018.

\bibitem{olgun2024read}
A.~Olgun, M.~Osseiran, A.~G. Yaglikci, Y.~C. Tugrul, H.~Luo, S.~Rhyner,
  B.~Salami, J.~Gomez~Luna, and O.~Mutlu, ``{Read Disturbance in High Bandwidth
  Memory: A Detailed Experimental Study on HBM2 DRAM Chips},'' in \emph{DSN},
  2024.

\bibitem{park2016experiments}
K.~Park, C.~Lim, D.~Yun, and S.~Baeg, ``{Experiments and Root Cause Analysis
  for Active-Precharge Hammering Fault in DDR3 SDRAM under 3xnm Technology},''
  \emph{Microelectronics Reliability}, 2016.

\bibitem{park2016statistical}
K.~Park, D.~Yun, and S.~Baeg, ``{Statistical Distributions of Row-Hammering
  Induced Failures in DDR3 Components},'' \emph{Microelectronics Reliability},
  2016.

\bibitem{pessl2016drama}
P.~Pessl, D.~Gruss, C.~Maurice, M.~Schwarz, and S.~Mangard, ``{DRAMA:
  Exploiting DRAM Addressing for Cross-CPU Attacks},'' in \emph{USENIX
  Security}, 2016.

\bibitem{poddebniak2018attacking}
D.~Poddebniak, J.~Somorovsky, S.~Schinzel, M.~Lochter, and P.~R{\"o}sler,
  ``{Attacking Deterministic Signature Schemes using Fault Attacks},'' in
  \emph{EuroS\&P}, 2018.

\bibitem{qiao2016new}
R.~Qiao and M.~Seaborn, ``{A New Approach for RowHammer Attacks},'' in
  \emph{HOST}, 2016.

\bibitem{rakin2022deepsteal}
A.~S. Rakin, M.~H.~I. Chowdhuryy, F.~Yao, and D.~Fan, ``{DeepSteal: Advanced
  Model Extractions Leveraging Efficient Weight Stealing in Memories},'' in
  \emph{S\&P}, 2022.

\bibitem{razavi2016flip}
K.~Razavi, B.~Gras, E.~Bosman, B.~Preneel, C.~Giuffrida, and H.~Bos, ``{Flip
  Feng Shui: Hammering a Needle in the Software Stack},'' in \emph{USENIX
  Security}, 2016.

\bibitem{roohi2022efficient}
A.~Roohi and S.~Angizi, ``{Efficient Targeted Bit-Flip Attack Against the Local
  Binary Pattern Network},'' in \emph{HOST}, 2022.

\bibitem{rowhammergithub}
{SAFARI Research Group}, ``{RowHammer --- GitHub Repository},''
  \url{https://github.com/CMU-SAFARI/rowhammer}, 2014.

\bibitem{ryu2017overcoming}
S.-W. Ryu, K.~Min, J.~Shin, H.~Kwon, D.~Nam, T.~Oh, T.-S. Jang, M.~Yoo, Y.~Kim,
  and S.~Hong, ``{Overcoming the Reliability Limitation in the Ultimately
  Scaled DRAM using Silicon Migration Technique by Hydrogen Annealing},'' in
  \emph{IEDM}, 2017.

\bibitem{seaborn2015exploiting}
\BIBentryALTinterwordspacing
M.~Seaborn and T.~Dullien, ``{Exploiting the DRAM Rowhammer Bug to Gain Kernel
  Privileges},'' Black Hat USA, 2015. [Online]. Available:
  \url{https://www.blackhat.com/docs/us-15/materials/us-15-Seaborn-Exploiting-The-DRAM-Rowhammer-Bug-To-Gain-Kernel-Privileges.pdf}
\BIBentrySTDinterwordspacing

\bibitem{shukla2026prowhammer}
M.~Shukla, S.~Roy, S.~Saha, and B.~Panda, ``{PRowhammer: Propagating Bit-flips
  from CPU to GPU},'' in \emph{ISCA}, 2026.

\bibitem{staudigl2022neurohammer}
F.~Staudigl, H.~Al~Indari, D.~Sch{\"o}n, D.~Sisejkovic, F.~Merchant, J.~M.
  Joseph, V.~Rana, S.~Menzel, and R.~Leupers, ``{NeuroHammer: Inducing
  Bit-Flips in Memristive Crossbar Memories},'' in \emph{DATE}, 2022.

\bibitem{tatar2018defeating}
A.~Tatar, C.~Giuffrida, H.~Bos, and K.~Razavi, ``{Defeating Software
  Mitigations Against Rowhammer: A Surgical Precision Hammer},'' in
  \emph{RAID}, 2018.

\bibitem{tatar2018throwhammer}
A.~Tatar, R.~K. Konoth, E.~Athanasopoulos, C.~Giuffrida, H.~Bos, and K.~Razavi,
  ``{Throwhammer: {Rowhammer} {Attacks} Over the {Network} and {Defenses}},''
  in \emph{{USENIX} {ATC}}, 2018.

\bibitem{tobah2022spechammer}
Y.~Tobah, A.~Kwong, I.~Kang, D.~Genkin, and K.~G. Shin, ``{SpecHammer:
  Combining Spectre and Rowhammer for New Speculative Attacks},'' in
  \emph{S\&P}, 2022.

\bibitem{tol2022toward}
\BIBentryALTinterwordspacing
M.~C. Tol, S.~Islam, A.~J. Adiletta, B.~Sunar, and Z.~Zhang, ``{Don't Knock!
  Rowhammer at the Backdoor of DNN Models},'' in \emph{DSN}, 2023. [Online].
  Available: \url{https://arxiv.org/abs/2110.07683v3}
\BIBentrySTDinterwordspacing

\bibitem{van2016drammer}
V.~van~der Veen, Y.~Fratantonio, M.~Lindorfer, D.~Gruss, C.~Maurice, G.~Vigna,
  H.~Bos, K.~Razavi, and C.~Giuffrida, ``{Drammer: Deterministic Rowhammer
  Attacks on Mobile Platforms},'' in \emph{CCS}, 2016.

\bibitem{van2018guardion}
V.~van~der Veen, M.~Lindorfer, Y.~Fratantonio, H.~Padmanabha~Pillai, G.~Vigna,
  C.~Kruegel, H.~Bos, and K.~Razavi, ``{GuardION: Practical Mitigation of
  DMA-Based Rowhammer Attacks on ARM},'' in \emph{{DIMVA}}, 2018.

\bibitem{walker2021ondramrowhammer}
A.~J. Walker, S.~Lee, and D.~Beery, ``{On DRAM RowHammer and the Physics of
  Insecurity},'' \emph{IEEE TED}, 2021.

\bibitem{wan2026geforge}
J.~Wan, Y.~Guo, Z.~Zhang, Z.~Li, D.~J. Tian, and Z.~Zhang, ``{GeForge:
  Hammering GDDR Memory to Forge GPU Page Tables for Fun and Profit},'' in
  \emph{S\&P}, 2026.

\bibitem{wang2022research}
J.~Wang, H.~Xu, C.~Xiao, L.~Zhang, and Y.~Zheng, ``{Research and Implementation
  of Rowhammer Attack Method based on Domestic NeoKylin Operating System},'' in
  \emph{ICFTIC}, 2022.

\bibitem{weissman2020jackhammer}
\BIBentryALTinterwordspacing
Z.~Weissman, T.~Tiemann, D.~Moghimi, E.~Custodio, T.~Eisenbarth, and B.~Sunar,
  ``{JackHammer: Efficient Rowhammer on Heterogeneous FPGA--CPU Platforms},''
  \emph{IACR Transactions on Cryptographic Hardware and Embedded Systems}, vol.
  2020, no.~3, pp. 169--195, 2020. [Online]. Available:
  \url{https://arxiv.org/abs/1912.11523}
\BIBentrySTDinterwordspacing

\bibitem{xiao2016one}
Y.~Xiao, X.~Zhang, Y.~Zhang, and R.~Teodorescu, ``{One Bit Flips, One Cloud
  Flops: Cross-VM Row Hammer Attacks and Privilege Escalation},'' in
  \emph{USENIX Security}, 2016.

\bibitem{yaglikci2022understanding}
A.~G. Ya{\u{g}}l{\i}k{\c{c}}{\i}, H.~Luo, G.~F. de~Oliveira, A.~Olgun,
  M.~Patel, J.~Park, H.~Hassan, J.~S. Kim, L.~Orosa, and O.~Mutlu,
  ``{Understanding RowHammer Under Reduced Wordline Voltage: An Experimental
  Study Using Real DRAM Devices},'' in \emph{DSN}, 2022, pp. 475--487.

\bibitem{yang2019trap}
T.~Yang and X.-W. Lin, ``{Trap-Assisted DRAM Row Hammer Effect},'' \emph{IEEE
  EDL}, 2019.

\bibitem{yang2022socially}
L.-H. Yang, S.-S. Huang, T.-L. Cheng, Y.-C. Kuo, and J.-J. Kuo,
  ``{Socially-Aware Collaborative Defense System against Bit-Flip Attack in
  Social Internet of Things and Its Online Assignment Optimization},'' in
  \emph{ICCCN}, 2022.

\bibitem{yao2020deephammer}
F.~Yao, A.~S. Rakin, and D.~Fan, ``{DeepHammer: Depleting the Intelligence of
  Deep Neural Networks Through Targeted Chain of Bit Flips},'' in \emph{USENIX
  Security}, 2020.

\bibitem{yuksel2025pudhammer}
I.~E. Yuksel, A.~Sood, A.~Olgun, O.~Canpolat, H.~Luo, N.~Bostanci,
  M.~Sadrosadati, A.~G. Ya{\u{g}}l{\i}k{\c{c}}{\i}, and O.~Mutlu, ``{PuDHammer:
  Experimental Analysis of Read Disturbance Effects of Processing-using-DRAM in
  Real DRAM Chips},'' in \emph{ISCA}, 2025.

\bibitem{yun2018study}
D.~Yun, M.~Park, C.~Lim, and S.~Baeg, ``{Study of TID Effects on One Row
  Hammering using Gamma in DDR4 SDRAMs},'' in \emph{IRPS}, 2018.

\bibitem{zhang2018triggering}
Z.~Zhang, Z.~Zhan, D.~Balasubramanian, X.~Koutsoukos, and G.~Karsai,
  ``{Triggering Rowhammer Hardware Faults on ARM: A Revisit},'' in
  \emph{ASHES}, 2018.

\bibitem{zhang2020pthammer}
Z.~Zhang, Y.~Cheng, D.~Liu, S.~Nepal, Z.~Wang, and Y.~Yarom, ``{PThammer:
  Cross-User-Kernel-Boundary Rowhammer through Implicit Accesses},'' in
  \emph{MICRO}, 2020.

\bibitem{zhang2022implicit}
Z.~Zhang, W.~He, Y.~Cheng, W.~Wang, Y.~Gao, D.~Liu, K.~Li, S.~Nepal, A.~Fu, and
  Y.~Zou, ``{Implicit Hammer: Cross-Privilege-Boundary Rowhammer through
  Implicit Accesses},'' \emph{IEEE TDSC}, 2023.

\bibitem{zheng2022trojvit}
\BIBentryALTinterwordspacing
M.~Zheng, Q.~Lou, and L.~Jiang, ``{TrojViT: Trojan Insertion in Vision
  Transformers},'' in \emph{CVPR}, 2023. [Online]. Available:
  \url{https://openaccess.thecvf.com/content/CVPR2023/html/Zheng_TrojViT_Trojan_Insertion_in_Vision_Transformers_CVPR_2023_paper.html}
\BIBentrySTDinterwordspacing

\bibitem{mutlu2018rowhammer}
O.~Mutlu, ``{RowHammer},''
  \url{https://people.inf.ethz.ch/omutlu/pub/onur-Rowhammer-TopPicksinHardwareEmbeddedSecurity-November-8-2018.pdf},
  2018, {Top Picks in Hardware and Embedded Security}.

\bibitem{yaglikci2021blockhammer}
A.~G. Ya{\u{g}}l{\i}k{\c{c}}{\i}, M.~Patel, J.~S. Kim, R.~Azizi, A.~Olgun,
  L.~Orosa, H.~Hassan, J.~Park, K.~Kanellopoulos, T.~Shahroodi, S.~Ghose, and
  O.~Mutlu, ``{BlockHammer: Preventing RowHammer at Low Cost by Blacklisting
  Rapidly-Accessed DRAM Rows},'' in \emph{HPCA}, 2021, pp. 345--358.

\bibitem{park2020graphene}
Y.~Park, W.~Kwon, E.~Lee, T.~J. Ham, J.~H. Ahn, and J.~W. Lee, ``{Graphene:
  Strong yet Lightweight Row Hammer Protection},'' in \emph{MICRO}, 2020.

\bibitem{qureshi2022hydra}
M.~Qureshi, A.~Rohan, G.~Saileshwar, and P.~J. Nair, ``{Hydra: Enabling
  Low-Overhead Mitigation of Row-Hammer at Ultra-Low Thresholds via Hybrid
  Tracking},'' in \emph{ISCA}, 2022.

\bibitem{olgun2024abacus}
A.~Olgun, Y.~C. Tugrul, N.~Bostanci, I.~E. Yuksel, H.~Luo, S.~Rhyner, A.~G.
  Yaglikci, G.~F. Oliveira, and O.~Mutlu, ``{ABACuS: All-Bank Activation
  Counters for Scalable and Low Overhead RowHammer Mitigation},'' in
  \emph{USENIX Security}, 2024.

\bibitem{canpolat2025chronus}
O.~Canpolat, A.~G. Yağlıkçı, G.~F. Oliveira, A.~Olgun, N.~Bostancı,
  {\.I}.~E. Yüksel, H.~Luo, O.~Ergin, and O.~Mutlu, ``{Chronus: Understanding
  and Securing the Cutting-Edge Industry Solutions to DRAM Read Disturbance},''
  in \emph{HPCA}, 2025.

\bibitem{fiedler2026memory}
C.~Fiedler, J.~Juffinger, S.~R. Neela, M.~Heckel, H.~Weissteiner, A.~G.
  Ya{\u{g}}l{\i}k{\c{c}}{\i}, F.~Adamsky, and D.~Gruss, ``{Memory Band-Aid: A
  Principled Rowhammer Defense-in-Depth},'' in \emph{NDSS}, 2026.

\bibitem{hajiabadi2025charm}
A.~Hajiabadi, M.~Marazzi, and K.~Razavi, ``{CHaRM: Checkpointed and Hashed
  Counters for Flexible and Efficient RowHammer Mitigation},'' in \emph{CCS},
  2025.

\bibitem{jedec2024ddr5}
{JEDEC}, \emph{{JESD79-5C: DDR5 SDRAM Standard}}, 2024.

\bibitem{kim2023how}
M.~J. Kim, M.~Wi, J.~Park, S.~Ko, J.~Choi, H.~Nam, N.~S. Kim, J.~H. Ahn, and
  E.~Lee, ``{{How to Kill the Second Bird with One ECC: The Pursuit of Row
  Hammer Resilient DRAM}},'' in \emph{MICRO}, 2023.

\bibitem{kim2025per}
J.~Kim, S.~Baek, M.~Wi, H.~Nam, M.~J. Kim, S.~Lee, K.~Sohn, and J.~H. Ahn,
  ``Per-row activation counting on real hardware: Demystifying performance
  overheads,'' \emph{CAL}, 2025.

\bibitem{saxena2024rubix}
A.~Saxena, S.~Mathur, and M.~Qureshi, ``{Rubix: Reducing the Overhead of Secure
  Rowhammer Mitigations via Randomized Line-to-Row Mapping},'' in
  \emph{ASPLOS}, 2024.

\bibitem{saxena2024start}
A.~Saxena and M.~Qureshi, ``{START: Scalable Tracking for any Rowhammer
  Threshold},'' in \emph{HPCA}, 2024.

\bibitem{stefanDRFM}
S.~Saroiu, ``{DRFM: Directed Refresh Management},''
  \url{https://stefan.t8k2.com/rh/DRFM/}, 2022.

\bibitem{taneja2026mirza}
H.~Taneja, A.~Hajiabadi, M.~Marazzi, K.~Razavi, and M.~Qureshi, ``{MIRZA:
  Efficiently Mitigating Rowhammer with Randomization and ALERT},'' in
  \emph{HPCA}, 2026.

\bibitem{wi2026rowarmor}
M.~Wi, Y.~Yoo, Y.~Kim, J.~Shin, J.~Kim, Y.~Ryu, S.~Gorgin, J.~H. Ahn, and
  J.~Kim, ``{RowArmor: Efficient and Comprehensive Protection Against DRAM
  Disturbance Attacks},'' in \emph{ASPLOS}, 2026.

\bibitem{woo2023rampart}
S.~C. Woo, W.~Elsasser, M.~Hamburg, E.~Linstadt, M.~R. Miller, T.~Song, and
  J.~Tringali, ``{{RAMPART: RowHammer Mitigation and Repair for Server Memory
  Systems}},'' in \emph{MEMSYS}, 2023.

\bibitem{wu2026apt}
R.~Wu, M.~Zhang, Y.~Zhou, C.~Xie, and F.~Wu, ``{APT: Securing Against DRAM Read
  Disturbance via Adaptive Probabilistic In-DRAM Trackers},'' in \emph{ASPLOS},
  2026.

\bibitem{AppleRefInc}
{Apple Inc.}, ``{About the Security Content of Mac EFI Security Update
  2015-001},'' \url{https://support.apple.com/en-us/HT204934}, Jun. 2015.

\bibitem{Nale2019}
W.~Nale, ``{Memory Controller That Filters a Count of Row Activate Commands
  Collectively Sent to a Set of Memory Banks},'' 2021, {U.S.\ Patent
  10,969,997}.

\bibitem{Qureshi2024MINT}
M.~Qureshi, S.~Qazi, and A.~Jaleel, ``{MINT: Securely Mitigating Rowhammer with
  a Minimalist in-DRAM Tracker},'' in \emph{MICRO}, 2024, pp. 899--914.

\bibitem{ajorpaz2022evax}
S.~M. Ajorpaz, D.~Moghimi, J.~N. Collins, G.~Pokam, N.~Abu-Ghazaleh, and
  D.~Tullsen, ``{EVAX: Towards a Practical, Pro-active \& Adaptive Architecture
  for High Performance \& Security},'' in \emph{MICRO}, 2022.

\bibitem{arikan2022processor}
K.~Ar{\i}kan, A.~Palumbo, L.~Cassano, P.~Reviriego, S.~Pontarelli, G.~Bianchi,
  O.~Ergin, and M.~Ottavi, ``{Processor security: Detecting microarchitectural
  attacks via count-min sketches},'' \emph{IEEE TVLSI}, 2022.

\bibitem{aweke2016anvil}
Z.~B. Aweke, S.~F. Yitbarek, R.~Qiao, R.~Das, M.~Hicks, Y.~Oren, and T.~Austin,
  ``{ANVIL: Software-Based Protection Against Next-Generation Rowhammer
  Attacks},'' in \emph{ASPLOS}, 2016.

\bibitem{bains-merged}
K.~Bains \emph{et~al.}, ``{Row Hammer Refresh Command},'' U.S.\ Patents
  9,117,544, 9,236,110, and 10,210,925, 2015--2019.

\bibitem{bains14c}
K.~S. Bains, J.~B. Halbert, S.~Sah, and Z.~Greenfield, ``{Method, Apparatus and
  System for Providing a Memory Refresh},'' US Patent: 9,030,903, 2015.

\bibitem{bains14d}
K.~Bains and J.~Halbert, ``{Distributed Row Hammer Tracking},'' U.S.\ Patent
  9,299,400, 2016.

\bibitem{bains2016row}
\BIBentryALTinterwordspacing
K.~S. Bains and J.~B. Halbert, ``{Row Hammer Monitoring Based on Stored Row
  Hammer Threshold Value},'' 2016, {U.S.}\ Patent 9,384,821. [Online].
  Available: \url{https://patents.google.com/patent/US9384821B2/en}
\BIBentrySTDinterwordspacing

\bibitem{bennett2021panopticon}
T.~Bennett, S.~Saroiu, A.~Wolman, and L.~Cojocar, ``{Panopticon: A Complete
  In-DRAM Rowhammer Mitigation},'' in \emph{DRAMSec}, 2021.

\bibitem{bock2019riprh}
C.~Bock, F.~Brasser, D.~Gens, C.~Liebchen, and A.-R. Sadeghi, ``{RIP-RH:
  Preventing Rowhammer-Based Inter-Process Attacks},'' in \emph{ASIACCS}, 2019.

\bibitem{bostanci2024comet}
F.~N. Bostancı, {\.I}.~E. Yüksel, A.~Olgun, K.~Kanellopoulos, Y.~C. Tuğrul,
  A.~G. Yağlıkçı, M.~Sadrosadati, and O.~Mutlu, ``{CoMeT:
  Count-Min-Sketch-Based Row Tracking to Mitigate RowHammer at Low Cost},'' in
  \emph{HPCA}, 2024, pp. 593--612.

\bibitem{canpolat2024breakhammer}
O.~Canpolat, A.~G. Ya{\u{g}}l{\i}k{\c{c}}{\i}, A.~Olgun, {\.I}.~E. Y{\"u}ksel,
  Y.~C. Tu{\u{g}}rul, K.~Kanellopoulos, O.~Ergin, and O.~Mutlu, ``{BreakHammer:
  Enhancing RowHammer Mitigations by Carefully Throttling Suspect Threads},''
  in \emph{MICRO}, 2024, pp. 915--934.

\bibitem{canpolat2024understanding}
O.~Canpolat, A.~G. Yağlıkçı, G.~F. Oliveira, A.~Olgun, O.~Ergin, and
  O.~Mutlu, ``{Understanding the Security Benefits and Overheads of Emerging
  Industry Solutions to DRAM Read Disturbance},'' in \emph{DRAMSec}, 2024.

\bibitem{devaux2021method}
F.~Devaux and R.~Ayrignac, ``{Method and Circuit for Protecting a DRAM Memory
  Device from the Row Hammer Effect},'' U.S.\ Patent 10,885,966, 2021.

\bibitem{di2023copy}
A.~Di~Dio, K.~Koning, H.~Bos, and C.~Giuffrida, ``{Copy-on-Flip: Hardening ECC
  Memory Against Rowhammer Attacks},'' in \emph{NDSS}, 2023.

\bibitem{enomoto2022efficient}
S.~Enomoto, H.~Kuzuno, and H.~Yamada, ``{Efficient Protection Mechanism for CPU
  Cache Flush Instruction Based Attacks},'' \emph{IEICE Transactions on
  Information and Systems}, 2022.

\bibitem{fakhrzadehgan2022safeguard}
A.~Fakhrzadehgan, Y.~N. Patt, P.~J. Nair, and M.~K. Qureshi, ``{SafeGuard:
  Reducing the Security Risk from Row-Hammer via Low-Cost Integrity
  Protection},'' in \emph{HPCA}, 2022.

\bibitem{france2022modeling}
L.~France, F.~Bruguier, M.~Mushtaq, D.~Novo, and P.~Benoit, ``{Modeling
  Rowhammer in the gem5 Simulator},'' in \emph{CHES (Poster)}, 2022.

\bibitem{france2022reducing}
L.~France, F.~Bruguier, D.~Novo, M.~Mushtaq, and P.~Benoit, ``{Reducing the
  Silicon Area Overhead of Counter-Based Rowhammer Mitigations},'' \emph{IEEE
  CAL}, 2024.

\bibitem{greenfield2012throttling}
Z.~Greenfield and T.~Levy, ``{Throttling Support for Row-Hammer Counters},''
  2016, {U.S.\ Patent 9,251,885}.

\bibitem{guha2022criticality}
K.~Guha and A.~Chakrabarti, ``{Criticality based Reliability from Rowhammer
  Attacks in Multi-User-Multi-FPGA Platform},'' in \emph{VLSID}, 2022.

\bibitem{han2021surround}
J.-W. Han, J.~Kim, D.~Beery, K.~D. Bozdag, P.~Cuevas, A.~Levi, I.~Tain,
  K.~Tran, A.~J. Walker, S.~V. Palayam, A.~Arreghini, A.~Furnémont, and
  M.~Meyyappan, ``{Surround Gate Transistor With Epitaxially Grown Si Pillar
  and Simulation Study on Soft Error and Rowhammer Tolerance for DRAM},''
  \emph{IEEE TED}, 2021.

\bibitem{hassan2024self}
H.~Hassan, A.~Olgun, A.~G. Ya{\u{g}}l{\i}k{\c{c}}{\i}, H.~Luo, and O.~Mutlu,
  ``{Self-Managing DRAM: A Low-Cost Framework for Enabling Autonomous and
  Efficient DRAM Maintenance Operations},'' in \emph{MICRO}, 2024.

\bibitem{hong2023dsac}
\BIBentryALTinterwordspacing
S.~Hong, D.~Kim, J.~Lee, R.~Oh, C.~Yoo, S.~Hwang, and J.~Lee, ``{DSAC: Low-Cost
  Rowhammer Mitigation Using In-DRAM Stochastic and Approximate Counting
  Algorithm},'' arXiv:2302.03591, 2023. [Online]. Available:
  \url{https://arxiv.org/abs/2302.03591}
\BIBentrySTDinterwordspacing

\bibitem{hp2015rowhammer}
{Hewlett-Packard Enterprise}, ``{HP Moonshot Component Pack Version
  2015.05.0},''
  \url{http://h17007.www1.hp.com/us/en/enterprise/servers/products/moonshot/component-pack/index.aspx},
  2015.

\bibitem{irazoqui2016mascat}
G.~Irazoqui, T.~Eisenbarth, and B.~Sunar, ``{MASCAT: Preventing
  Microarchitectural Attacks Before Distribution},'' in \emph{CODASPY}, 2018.

\bibitem{jaleel2024pride}
A.~Jaleel, G.~Saileshwar, S.~W. Keckler, and M.~Qureshi, ``{PrIDE: Achieving
  Secure Rowhammer Mitigation with Low-Cost In-DRAM Trackers},'' in
  \emph{ISCA}.\hskip 1em plus 0.5em minus 0.4em\relax IEEE, 2024, pp.
  1157--1172.

\bibitem{joardar2022learning}
B.~K. Joardar, T.~K. Bletsch, and K.~Chakrabarty, ``{Learning to Mitigate
  RowHammer Attacks},'' in \emph{DATE}, 2022.

\bibitem{joardar2022machine}
------, ``{Machine Learning-based Rowhammer Mitigation},'' \emph{TCAD}, 2023.

\bibitem{juffinger2023csi}
J.~Juffinger, L.~Lamster, A.~Kogler, M.~Eichlseder, M.~Lipp, and D.~Gruss,
  ``{CSI: Rowhammer--Cryptographic Security and Integrity against Rowhammer},''
  in \emph{S\&P}, 2023.

\bibitem{kakolyris2026columnkeeper}
A.~K. Kakolyris, F.~N. Bostanci, A.~Olgun, I.~E. Yuksel, H.~Songara, K.~M.
  Sgouras, U.~Baser, K.~Kanellopoulos, A.~G. Yaglikci, and O.~Mutlu,
  ``{ColumnKeeper: Efficient Solutions to the ColumnDisturb Vulnerability in
  DRAM-based Systems},'' in \emph{ISCA}, 2026.

\bibitem{kang2020cattwo}
I.~Kang, E.~Lee, and J.~H. Ahn, ``{CAT-TWO: Counter-Based Adaptive Tree, Time
  Window Optimized for {DRAM} Row-Hammer Prevention},'' \emph{{IEEE} Access},
  2020.

\bibitem{kim2015architectural}
D.-H. Kim, P.~J. Nair, and M.~K. Qureshi, ``{Architectural Support for
  Mitigating Row Hammering in DRAM Memories},'' \emph{CAL}, 2015.

\bibitem{kim2021hammerfilter}
K.~Kim, J.~Woo, J.~Kim, and K.-S. Chung, ``{HammerFilter: Robust Protection and
  Low Hardware Overhead Method for RowHammer},'' in \emph{ICCD}, 2021.

\bibitem{kim2022mithril}
M.~J. Kim, J.~Park, Y.~Park, W.~Doh, N.~Kim, T.~J. Ham, J.~W. Lee, and J.~H.
  Ahn, ``{Mithril: Cooperative Row Hammer Protection on Commodity DRAM
  Leveraging Managed Refresh},'' in \emph{HPCA}, 2022.

\bibitem{kim2023ddr5}
W.~Kim, C.~Jung, S.~Yoo, D.~Hong, J.~Hwang, J.~Yoon, O.~Jung, J.~Choi, S.~Hyun,
  M.~Kang, S.~Lee, D.~Kim, S.~Ku, D.~Choi, N.~Joo, S.~Yoon, J.~Noh, B.~Go,
  C.~Kim, S.~Hwang, M.~Hwang, S.-M. Yi, H.~Kim, S.~Heo, Y.~Jang, K.~Jang,
  S.~Chu, Y.~Oh, K.~Kim, J.~Kim, S.~Kim, J.~Hwang, S.~Park, J.~Lee, I.~Jeong,
  J.~Cho, and J.~Kim, ``{A 1.1V 16Gb DDR5 DRAM with Probabilistic-Aggressor
  Tracking, Refresh-Management Functionality, Per-Row Hammer Tracking, a
  Multi-Step Precharge, and Core-Bias Modulation for Security and Reliability
  Enhancement},'' in \emph{ISSCC}.\hskip 1em plus 0.5em minus 0.4em\relax IEEE,
  2023, pp. 1--3.

\bibitem{konoth2018zebram}
R.~K. Konoth, M.~Oliverio, A.~Tatar, D.~Andriesse, H.~Bos, C.~Giuffrida, and
  K.~Razavi, ``{ZebRAM: Comprehensive and Compatible Software Protection
  Against Rowhammer Attacks},'' in \emph{OSDI}, 2018.

\bibitem{lee2019twice}
E.~Lee, I.~Kang, S.~Lee, G.~{Edward Suh}, and J.~{Ho Ahn}, ``{TWiCe: Preventing
  Row-Hammering by Exploiting Time Window Counters},'' in \emph{ISCA}, 2019.

\bibitem{lee2021cryoguard}
G.-H. Lee, S.~Na, I.~Byun, D.~Min, and J.~Kim, ``{CryoGuard: A Near
  Refresh-Free Robust DRAM Design for Cryogenic Computing},'' in \emph{ISCA},
  2021.

\bibitem{lenovo2015rowhammer}
{Lenovo Group Ltd.}, ``{Row Hammer Privilege Escalation},''
  \url{https://support.lenovo.com/us/en/product_security/row_hammer}, 2015.

\bibitem{lin2015handling}
J.~Lin and M.~Garrett, ``{Handling Maximum Activation Count Limit and Target
  Row Refresh in DDR4 SDRAM},'' 2017, {U.S.\ Patent 9,589,606}.

\bibitem{lin2025cnc}
C.~S. Lin, J.~Woo, P.~J. Nair, and G.~Saileshwar, ``{CnC-PRAC: Coalesce, not
  Cache, Per Row Activation Counts for an Efficient in-DRAM Rowhammer
  Mitigation},'' in \emph{DRAMSec}, 2025.

\bibitem{loughlin2021stop}
K.~Loughlin, S.~Saroiu, A.~Wolman, and B.~Kasikci, ``{Stop! Hammer Time:
  Rethinking Our Approach to Rowhammer Mitigations},'' in \emph{HotOS}, 2021.

\bibitem{loughlin2022moesiprime}
K.~Loughlin, S.~Saroiu, A.~Wolman, Y.~A. Manerkar, and B.~Kasikci,
  ``{MOESI-Prime: Preventing Coherence-Induced Hammering in Commodity
  Workloads},'' in \emph{ISCA}, 2022.

\bibitem{lu2025counterpoint}
S.-L. Lu, J.~Woo, and P.~J. Nair, ``{Counterpoint: One-Hot Counting for
  PRAC-Based RowHammer Mitigation},'' in \emph{DRAMSec}, 2025.

\bibitem{manzhosov2022revisiting}
E.~Manzhosov, A.~Hastings, M.~Pancholi, R.~Piersma, M.~T.~I. Ziad, and
  S.~Sethumadhavan, ``{Revisiting Residue Codes for Modern Memories},'' in
  \emph{MICRO}, 2022.

\bibitem{marazzi2022protrr}
M.~Marazzi, P.~Jattke, F.~Solt, and K.~Razavi, ``{ProTRR}: {Principled} yet
  {Optimal} {In-DRAM} {Target Row Refresh},'' in \emph{{S\&P}}, 2022.

\bibitem{marazzi2023rega}
M.~Marazzi, F.~Solt, P.~Jattke, K.~Takashi, and K.~Razavi, ``{{REGA}: Scalable
  Rowhammer Mitigation with Refresh-Generating Activations},'' in
  \emph{{S\&P}}, 2023, pp. 1684--1701.

\bibitem{naseredini2022alarm}
A.~Naseredini, M.~Berger, M.~Sammartino, and S.~Xiong, ``{ALARM: Active
  LeArning of Rowhammer Mitigations},'' in \emph{HASP}, 2022.

\bibitem{park2022rowhammer}
J.~H. Park, S.~Y. Kim, D.~Y. Kim, G.~Kim, J.~W. Park, S.~Yoo, Y.-W. Lee, and
  M.~J. Lee, ``{RowHammer Reduction Using a Buried Insulator in a Buried
  Channel Array Transistor},'' \emph{IEEE Transactions on Electron Devices},
  2022.

\bibitem{qazi2025drfm}
S.~Qazi and M.~Qureshi, ``{DRFM and the Art of Rowhammer Sampling},'' in
  \emph{DRAMSec}, 2025.

\bibitem{qureshi2024impress}
A.~Saxena, A.~Jaleel, and M.~Qureshi, ``{ImPress: Securing DRAM Against
  Data-Disturbance Errors via Implicit Row-Press Mitigation},'' in
  \emph{MICRO}, 2024.

\bibitem{qureshi2025autorfm}
M.~Qureshi, ``{AutoRFM: Scaling Low-Cost in-DRAM Trackers to Ultra-Low
  Rowhammer Thresholds},'' in \emph{HPCA}, 2025.

\bibitem{qureshi2025moat}
M.~Qureshi and S.~Qazi, ``{MOAT: Securely Mitigating Rowhammer with Per-Row
  Activation Counters},'' in \emph{ASPLOS}.\hskip 1em plus 0.5em minus
  0.4em\relax Association for Computing Machinery, 2025, pp. 698--714.

\bibitem{qureshi2026salt}
M.~K. Qureshi, ``{SALT: Track-and-Mitigate Subarrays, Not Rows, for
  Blast-Radius-Free Rowhammer Defense},'' in \emph{HPCA}, 2026.

\bibitem{saileshwar2022randomized}
G.~Saileshwar, B.~Wang, M.~Qureshi, and P.~J. Nair, ``{Randomized Row-Swap:
  Mitigating Row Hammer by Breaking Spatial Correlation Between Aggressor and
  Victim Rows},'' in \emph{ASPLOS}, 2022.

\bibitem{saroiu2022configure}
S.~Saroiu and A.~Wolman, ``{How to Configure Row-Sampling-Based Rowhammer
  Defenses},'' in \emph{DRAMSec}, 2022.

\bibitem{saroiu2022price}
S.~Saroiu, A.~Wolman, and L.~Cojocar, ``{The Price of Secrecy: How Hiding
  Internal DRAM Topologies Hurts Rowhammer Defenses},'' in \emph{IRPS}, 2022.

\bibitem{saroiu2024ddr5}
S.~Saroiu, ``{DDR5 Spec Update Has All It Needs to End Rowhammer: Will It?}''
  \url{https://stefan.t8k2.com/rh/PRAC/index.html}, 2024.

\bibitem{saxena2022aqua}
A.~Saxena, G.~Saileshwar, P.~J. Nair, and M.~Qureshi, ``{AQUA: Scalable
  Rowhammer Mitigation by Quarantining Aggressor Rows at Runtime},'' in
  \emph{MICRO}, 2022.

\bibitem{saxena2023pt}
A.~Saxena, G.~Saileshwar, J.~Juffinger, A.~Kogler, D.~Gruss, and M.~Qureshi,
  ``{PT-Guard: Integrity-Protected Page Tables to Defend Against Breakthrough
  Rowhammer Attacks},'' in \emph{DSN}, 2023.

\bibitem{seyedzadeh2017counter}
S.~M. Seyedzadeh, A.~K. Jones, and R.~Melhem, ``{Counter-based Tree Structure
  for Row Hammering Mitigation in DRAM},'' \emph{IEEE CAL}, 2017.

\bibitem{seyedzadeh2018cbt}
S.~M. {Seyedzadeh}, A.~K. {Jones}, and R.~{Melhem}, ``{Mitigating Wordline
  Crosstalk Using Adaptive Trees of Counters},'' in \emph{ISCA}, 2018.

\bibitem{son2017making}
M.~Son, H.~Park, J.~Ahn, and S.~Yoo, ``{Making DRAM Stronger Against Row
  Hammering},'' in \emph{DAC}, 2017.

\bibitem{taneja2025dream}
H.~Taneja and M.~Qureshi, ``{DREAM: Enabling Low-Overhead Rowhammer Mitigation
  via Directed Refresh Management},'' in \emph{ISCA}, 2025.

\bibitem{tomita2022extracting}
C.~Tomita, M.~Takita, K.~Fukushima, Y.~Nakano, Y.~Shiraishi, and M.~Morii,
  ``{Extracting the Secrets of OpenSSL with RAMBleed},'' \emph{Sensors}, 2022.

\bibitem{tugrul2025understanding}
Y.~C. Tugrul, A.~G. Yaglikci, I.~E. Yuksel, A.~Olgun, O.~Canpolat, N.~Bostanci,
  M.~Sadrosadati, O.~Ergin, and O.~Mutlu, ``{Understanding RowHammer Under
  Reduced Refresh Latency: Experimental Analysis of Real DRAM Chips and
  Implications on Future Solutions},'' in \emph{HPCA}, 2025.

\bibitem{vig2018rapid}
S.~Vig, S.~Bhattacharya, D.~Mukhopadhyay, and S.-K. Lam, ``{Rapid Detection of
  Rowhammer Attacks Using Dynamic Skewed Hash Tree},'' in \emph{HASP}, 2018.

\bibitem{vittal2025mopac}
S.~Vittal, S.~Qazi, P.~Das, and M.~Qureshi, ``{MoPAC: Efficiently Mitigating
  Rowhammer with Probabilistic Activation Counting},'' in \emph{ISCA}, 2025.

\bibitem{wang2021discreet}
Y.~Wang, Y.~Liu, P.~Wu, and Z.~Zhang, ``{Discreet-PARA: Rowhammer Defense with
  Low Cost and High Efficiency},'' in \emph{ICCD}, 2021, pp. 433--441.

\bibitem{wi2023shadow}
M.~Wi, J.~Park, S.~Ko, M.~J. Kim, N.~S. Kim, E.~Lee, and J.~H. Ahn, ``{SHADOW:
  Preventing Row Hammer in DRAM with Intra-Subarray Row Shuffling},'' in
  \emph{HPCA}, 2023.

\bibitem{woo2023scalable}
J.~Woo, G.~Saileshwar, and P.~J. Nair, ``{Scalable and Secure Row-Swap:
  Efficient and Safe Row Hammer Mitigation in Memory Systems},'' in
  \emph{HPCA}, 2023.

\bibitem{woo2025dapper}
J.~Woo and P.~J. Nair, ``{DAPPER: A Performance-Attack-Resilient Tracker for
  RowHammer Defense},'' in \emph{HPCA}, 2025.

\bibitem{woo2025qprac}
J.~Woo, S.~C. Lin, P.~J. Nair, A.~Jaleel, and G.~Saileshwar, ``{QPRAC: Towards
  Secure and Practical PRAC-based Rowhammer Mitigation using Priority
  Queues},'' in \emph{HPCA}, 2025.

\bibitem{yaglikci2021security}
\BIBentryALTinterwordspacing
A.~G. Ya{\u{g}}l{\i}k{\c{c}}{\i}, J.~S. Kim, F.~Devaux, and O.~Mutlu,
  ``{Security Analysis of the Silver Bullet Technique for RowHammer
  Prevention},'' arXiv:2106.07084, 2021. [Online]. Available:
  \url{https://arxiv.org/abs/2106.07084}
\BIBentrySTDinterwordspacing

\bibitem{yaglikci2022hira}
A.~G. Yağlıkçı, A.~Olgun, M.~Patel, H.~Luo, H.~Hassan, L.~Orosa, O.~Ergin,
  and O.~Mutlu, ``{HiRA: Hidden Row Activation for Reducing Refresh Latency of
  Off-the-Shelf DRAM Chips},'' in \emph{MICRO}, 2022.

\bibitem{yaglikci2024spatial}
A.~G. Yağlıkçı, Y.~C. Tuğrul, G.~F. Oliveira, {\.I}.~E. Yüksel, A.~Olgun,
  H.~Luo, and O.~Mutlu, ``{Spatial Variation-Aware Read Disturbance Defenses:
  Experimental Analysis of Real DRAM Chips and Implications on Future
  Solutions},'' in \emph{HPCA}, 2024.

\bibitem{you2019mrloc}
J.~M. You and J.-S. Yang, ``{MRLoc: Mitigating Row-Hammering Based on Memory
  Locality},'' in \emph{DAC}, 2019.

\bibitem{zhang2020leveraging}
Z.~Zhang, Z.~Zhan, D.~Balasubramanian, B.~Li, P.~Volgyesi, and X.~Koutsoukos,
  ``{Leveraging EM Side-Channel Information to Detect Rowhammer Attacks},'' in
  \emph{S\&P}, 2020.

\bibitem{zhang2022softtrr}
Z.~Zhang, Y.~Cheng, M.~Wang, W.~He, W.~Wang, S.~Nepal, Y.~Gao, K.~Li, Z.~Wang,
  and C.~Wu, ``{SoftTRR: Protect Page Tables against Rowhammer Attacks using
  Software-only Target Row Refresh},'' in \emph{USENIX ATC}, 2022.

\bibitem{zhou2022ltpim}
R.~Zhou, S.~Tabrizchi, A.~Roohi, and S.~Angizi, ``{LT-PIM: An LUT-based
  Processing-in-DRAM Architecture with RowHammer Self-Tracking},'' \emph{IEEE
  CAL}, 2022.

\bibitem{zhou2023dnndefender}
R.~Zhou, S.~Ahmed, A.~S. Rakin, and S.~Angizi, ``{DNN-Defender: A
  Victim-Focused In-DRAM Defense Mechanism for Taming Adversarial Weight Attack
  on DNNs},'' in \emph{DAC}, 2024.

\bibitem{olgun2025variable}
A.~Olgun, F.~N. Bostanci, I.~E. Yuksel, O.~Canpolat, H.~Luo, G.~F. Oliveira,
  A.~G. Yaglikci, M.~Patel, and O.~Mutlu, ``{Variable Read Disturbance: An
  Experimental Analysis of Temporal Variation in DRAM Read Disturbance},'' in
  \emph{HPCA}, 2025.

\bibitem{luo2025revisiting}
H.~Luo, {\.I}.~E. Y{\"u}ksel, A.~Olgun, A.~G. Ya{\u{g}}l{\i}k{\c{c}}{\i}, and
  O.~Mutlu, ``{Revisiting DRAM Read Disturbance: Identifying Inconsistencies
  Between Experimental Characterization and Device-Level Studies},'' in
  \emph{VTS}, 2025.

\bibitem{vts25github}
{SAFARI Research Group}, ``{ReadDisturbanceVTS25},''
  \url{https://github.com/CMU-SAFARI/ReadDisturbanceVTS25}, 2025.

\bibitem{zhou2023double}
L.~Zhou, J.~Li, Z.~Qiao, P.~Ren, Z.~Sun, J.~Wang, B.~Wu, Z.~Ji, R.~Wang,
  K.~Cao, and R.~Huang, ``{Double-sided Row Hammer Effect in Sub-20 nm DRAM:
  Physical Mechanism, Key Features and Mitigation},'' in \emph{IRPS}, 2023.

\bibitem{Zhou2024Understanding}
L.~Zhou, J.~Li, P.~Ren, S.~Ye, D.~Wang, Z.~Qiao, and Z.~Ji, ``{Understanding
  the Physical Mechanism of RowPress at the Device-Level in Sub-20 nm DRAM},''
  in \emph{IRPS}, 2024, pp. 1--6.

\bibitem{thomas2023hammulator}
F.~Thomas, L.~Gerlach, and M.~Schwarz, ``{Hammulator: Simulate Now -- Exploit
  Later},'' in \emph{DRAMSec}, 2023.

\bibitem{goswami2026hammersimsystemleveltoolmodel}
K.~Goswami, A.~Akram, H.~Venugopalan, and J.~Lowe-Power, ``{HammerSim: A
  System-Level Tool to Model RowHammer},'' in \emph{IISWC}, 2026.

\bibitem{luo2026demystifying}
\BIBentryALTinterwordspacing
H.~Luo, L.~Zhou, A.~Olgun, {\.I}.~E. Yüksel, N.~Bostanci, Z.~Ji, X.~Wu, and
  O.~Mutlu, ``{Demystifying DRAM Read Disturbance: Bridging the Gap Between
  Experimental Characterization and Device-Level Modeling of RowHammer and
  RowPress Phenomena},'' arXiv:2607.28233 [cs.AR], 2026. [Online]. Available:
  \url{https://arxiv.org/abs/2607.28233}
\BIBentrySTDinterwordspacing

\bibitem{bai2026unified}
\BIBentryALTinterwordspacing
Z.~Bai, H.~Luo, X.~Wang, X.~Li, Y.~Jiang, X.~Huang, G.~Wang, H.~Meng, A.~Olgun,
  I.~E. Yuksel, F.~N. Bostancı, J.~Wang, and O.~Mutlu, ``{A Unified
  Device-Level Physical Framework for DRAM Read Disturbance},''
  arXiv:2611.17634 [cond-mat.mes-hall], 2026. [Online]. Available:
  \url{https://arxiv.org/abs/2611.17634}
\BIBentrySTDinterwordspacing

\bibitem{Gautam2020MitigatingPassing}
S.~K. Gautam, S.~K. Manhas, A.~Kumar, and M.~Pakala, ``{Mitigating the Passing
  Word Line Induced Soft Errors in Saddle Fin DRAM},'' \emph{IEEE TED}, 2020.

\bibitem{Zhou2024Unveiling}
L.~Zhou, S.~Ye, R.~Wang, and Z.~Ji, ``{Unveiling RowPress in Sub-20 nm DRAM
  Through Comparative Analysis With Row Hammer: From Leakage Mechanisms to Key
  Features},'' \emph{IEEE Transactions on Electron Devices}, vol.~71, no.~8,
  pp. 4677--4684, 2024.

\bibitem{Jie2024Understanding}
J.~Li, L.~Zhou, S.~Ye, Z.~Qiao, and Z.~Ji, ``{Understanding the Competitive
  Interaction in Leakage Mechanisms for Effective Row Hammer Mitigation in
  Sub-20 nm DRAM},'' \emph{IEEE Electron Device Letters}, 2024.

\bibitem{kim2012case}
Y.~Kim, V.~Seshadri, D.~Lee, J.~Liu, and O.~Mutlu, ``{A Case for Exploiting
  Subarray-Level Parallelism (SALP) in DRAM},'' in \emph{ISCA}, 2012.

\bibitem{nam2024dramscope}
H.~Nam, S.~Baek, M.~Wi, M.~J. Kim, J.~Park, C.~Song, N.~S. Kim, and J.~H. Ahn,
  ``{DRAMScope: Uncovering DRAM Microarchitecture and Characteristics by
  Issuing Memory Commands},'' in \emph{ISCA}, 2024.

\bibitem{park2014active}
K.~Park, S.~Baeg, S.~Wen, and R.~Wong, ``{Active-Precharge Hammering on a
  Row-Induced Failure in DDR3 SDRAMs Under 3x nm Technology},'' in \emph{IIRW},
  2014.

\bibitem{yang2016suppression}
C.~Yang, C.~K. Wei, Y.~J. Chang, T.~C. Wu, H.~P. Chen, and C.~S. Lai,
  ``{Suppression of RowHammer Effect by Doping Profile Modification in
  Saddle-Fin Array Devices for Sub-30-nm DRAM Technology},'' \emph{TDMR}, 2016.

\bibitem{gautam2019rowhammering}
S.~K. Gautam, S.~K. Manhas, A.~Kumar, M.~Pakala, and E.~Yieh, ``{Row Hammering
  Mitigation Using Metal Nanowire in Saddle Fin DRAM},'' \emph{IEEE TED}, 2019.

\bibitem{france2021implementing}
L.~France, F.~Bruguier, M.~Mushtaq, D.~Novo, and P.~Benoit, ``{Implementing
  Rowhammer Memory Corruption in the gem5 Simulator},'' in \emph{RSP}, 2021.

\bibitem{demirel2025circuit}
E.~D. Demirel, E.~Afacan, and G.~D{\"u}ndar, ``{Circuit-Level Modeling and
  Simulation of Read Disturbance Phenomena: RowHammer and RowPress},'' in
  \emph{SMACD}, 2025.

\bibitem{jiang2021quantifying}
Y.~Jiang, H.~Zhu, D.~Sullivan, X.~Guo, X.~Zhang, and Y.~Jin, ``{Quantifying
  RowHammer Vulnerability for DRAM Security},'' in \emph{DAC}, 2021.

\bibitem{hu2026quantifying}
Z.~Hu, H.~Fei, P.~Gope, J.~Miskelly, O.~Millwood, and B.~Sikdar, ``{Quantifying
  Memory Cells Vulnerability for DRAM Security},'' in \emph{EuroS\&P}, 2026,
  arXiv:2603.18549.

\bibitem{luo2023ramulator2}
\BIBentryALTinterwordspacing
H.~Luo, Y.~C. Tu\u{g}rul, F.~N. Bostancı, A.~Olgun, A.~G.
  Ya\u{g}l{\i}k\c{c}{\i}, and O.~Mutlu, ``{Ramulator 2.0: A Modern, Modular,
  and Extensible DRAM Simulator},'' \emph{IEEE Computer Architecture Letters},
  vol.~23, no.~1, pp. 112--116, 2024. [Online]. Available:
  \url{https://arxiv.org/abs/2308.11030}
\BIBentrySTDinterwordspacing

\bibitem{olgun2023dram}
A.~Olgun, H.~Hassan, A.~G. Ya{\u{g}}l{\i}k{\c{c}}{\i}, Y.~C. Tu{\u{g}}rul,
  L.~Orosa, H.~Luo, M.~Patel, O.~Ergin, and O.~Mutlu, ``{DRAM Bender: An
  Extensible and Versatile FPGA-based Infrastructure to Easily Test
  State-of-the-art DRAM Chips},'' \emph{IEEE TCAD}, 2023.

\bibitem{safari-drambender}
{SAFARI Research Group}, ``{DRAM Bender --- GitHub Repository},''
  \url{https://github.com/CMU-SAFARI/DRAM-Bender}, 2022.

\bibitem{liu2013experimental}
J.~Liu, B.~Jaiyen, Y.~Kim, C.~Wilkerson, and O.~Mutlu, ``{An Experimental Study
  of Data Retention Behavior in Modern DRAM Devices},'' in \emph{ISCA}, 2013.

\bibitem{binkert2011gem5}
\BIBentryALTinterwordspacing
N.~Binkert, B.~Beckmann, G.~Black, S.~K. Reinhardt, A.~Saidi, A.~Basu,
  J.~Hestness, D.~R. Hower, T.~Krishna, S.~Sardashti, R.~Sen, K.~Sewell,
  M.~Shoaib, N.~Vaish, M.~D. Hill, and D.~A. Wood, ``{The Gem5 Simulator},''
  \emph{SIGARCH Comput. Archit. News}, vol.~39, no.~2, pp. 1--7, Aug. 2011.
  [Online]. Available: \url{https://doi.org/10.1145/2024716.2024718}
\BIBentrySTDinterwordspacing

\bibitem{luo2026ramulator21}
\BIBentryALTinterwordspacing
H.~Luo, F.~N. Bostanc{\i}, A.~Olgun, M.~Makeenkova, Z.~Malik, I.~Akdeniz, and
  O.~Mutlu, ``{Ramulator 2.1: A Composable Memory System Simulator for Modern
  DRAM Systems},'' {arXiv:2606.13844}, 2026. [Online]. Available:
  \url{https://arxiv.org/abs/2606.13844}
\BIBentrySTDinterwordspacing

\bibitem{ramulator2github}
{SAFARI Research Group}, ``{Ramulator V2.0},''
  \url{https://github.com/CMU-SAFARI/ramulator2}, 2023.

\bibitem{bruce2021enabling}
B.~R. Bruce, A.~Akram, H.~Nguyen, K.~Roarty, M.~Samani, M.~Friborz, T.~Reddy,
  M.~D. Sinclair, and J.~Lowe-Power, ``{Enabling Reproducible and Agile
  Full-System Simulation},'' in \emph{ISPASS}, 2021, pp. 183--193.

\bibitem{gem5resources}
``{gem5 Resources},'' \url{https://resources.gem5.org}, resources
  x86-linux-kernel-6.8.0-52-generic (version 1.0.0) and x86-ubuntu-24.04-img
  (version 4.0.0).

\bibitem{jedec2020ddr4}
{JEDEC}, \emph{{JESD79-4C: DDR4 SDRAM Standard}}, 2020.

\bibitem{khan2014efficacy}
S.~Khan, D.~Lee, Y.~Kim, A.~R. Alameldeen, C.~Wilkerson, and O.~Mutlu, ``{The
  Efficacy of Error Mitigation Techniques for {DRAM} Retention Failures: A
  Comparative Experimental Study},'' in \emph{SIGMETRICS}, 2014.

\bibitem{patel2017reaper}
M.~Patel, J.~S. Kim, and O.~Mutlu, ``{The Reach Profiler (REAPER): Enabling the
  Mitigation of DRAM Retention Failures via Profiling at Aggressive
  Conditions},'' in \emph{ISCA}, 2017.

\bibitem{chang2016understanding}
K.~K. Chang, A.~Kashyap, H.~Hassan, S.~Ghose, K.~Hsieh, D.~Lee, T.~Li,
  G.~Pekhimenko, S.~Khan, and O.~Mutlu, ``{Understanding Latency Variation in
  Modern DRAM Chips: Experimental Characterization, Analysis, and
  Optimization},'' in \emph{SIGMETRICS}, 2016.

\bibitem{lee2017design}
D.~Lee, S.~Khan, L.~Subramanian, S.~Ghose, R.~Ausavarungnirun, G.~Pekhimenko,
  V.~Seshadri, and O.~Mutlu, ``{Design-Induced Latency Variation in Modern DRAM
  Chips: Characterization, Analysis, and Latency Reduction Mechanisms},'' in
  \emph{SIGMETRICS}, 2017.

\bibitem{kim2018solar}
J.~S. Kim, M.~Patel, H.~Hassan, and O.~Mutlu, ``{Solar-DRAM: Reducing DRAM
  Access Latency by Exploiting the Variation in Local Bitlines},'' in
  \emph{ICCD}, 2018.

\bibitem{chang2017understanding}
K.~K. Chang, A.~G. Yağlıkçı, S.~Ghose, A.~Agrawal, N.~Chatterjee,
  A.~Kashyap, D.~Lee, M.~O'Connor, H.~Hassan, and O.~Mutlu, ``{Understanding
  Reduced-Voltage Operation in Modern DRAM Devices: Experimental
  Characterization, Analysis, and Mechanisms},'' in \emph{SIGMETRICS}, 2017.

\bibitem{wang2026scaledisturb}
J.~Wang, H.~Luo, A.~Olgun, I.~E. Y{\"u}ksel, A.~G. Ya{\u{g}}l{\i}k{\c{c}}{\i},
  Y.~Liang, F.~N. Bostanci, M.~Sadrosadati, and O.~Mutlu, ``{ScaleDisturb:
  Exploiting Temporal Asymmetry to Amplify Read Disturbance in Modern DRAM
  Chips},'' in \emph{DSN}, 2026.

\bibitem{lim2018study}
C.~Lim, K.~Park, G.~Bak, D.~Yun, M.~Park, S.~Baeg, S.-J. Wen, and R.~Wong,
  ``{Study of Proton Radiation Effect to Row Hammer Fault in DDR4 SDRAMs},''
  \emph{Microelectronics Reliability}, 2018.

\bibitem{baeg2022estimation}
S.~Baeg, D.~Yun, M.~Chun, and S.-J. Wen, ``{Estimation of the Trap Energy
  Characteristics of Row Hammer-Affected Cells in Gamma-Irradiated DDR4
  DRAM},'' \emph{IEEE Transactions on Nuclear Science}, 2022.

\bibitem{lang2023blaster}
Z.~Lang, P.~Jattke, M.~Marazzi, and K.~Razavi, ``{{BLASTER}: Characterizing the
  Blast Radius of Rowhammer},'' in \emph{DRAMSec}, 2023.

\bibitem{he2023whistleblower}
W.~He, Z.~Zhang, Y.~Cheng, W.~Wang, W.~Song, Y.~Gao, Q.~Zhang, K.~Li, D.~Liu,
  and S.~Nepal, ``{WhistleBlower: A System-level Empirical Study on
  RowHammer},'' \emph{IEEE Transactions on Computers}, 2025.

\bibitem{yuksel2025columndisturb}
{\.I}.~E. Yüksel, A.~Olgun, F.~N. Bostancı, H.~Luo, A.~G. Yağlıkçı, and
  O.~Mutlu, ``{ColumnDisturb: Understanding Column-based Read Disturbance in
  Real DRAM Chips and Implications for Future Systems},'' in \emph{MICRO},
  2025.

\bibitem{olgun2023hbm}
A.~Olgun, M.~Osseiran, A.~G. Yaglikci, Y.~C. Tugrul, H.~Luo, S.~Rhyner,
  B.~Salami, J.~Gomez~Luna, and O.~Mutlu, ``{An Experimental Analysis of
  RowHammer in HBM2 DRAM Chips},'' in \emph{DSN Disrupt}, 2023.

\bibitem{nam2023xray}
H.~Nam, S.~Baek, M.~Wi, M.~J. Kim, J.~Park, C.~Song, N.~S. Kim, and J.~H. Ahn,
  ``{X-ray: Discovering DRAM Internal Structure and Error Characteristics by
  Issuing Memory Commands},'' \emph{IEEE CAL}, 2023.

\bibitem{marazzi2024hifidram}
M.~Marazzi, T.~Sachsenweger, F.~Solt, P.~Zeng, T.~Kubo, M.~Yarema, and
  K.~Razavi, ``{HiFi-DRAM: Enabling High-fidelity DRAM Research by Uncovering
  Sense Amplifiers with IC Imaging},'' in \emph{ISCA}, 2024.

\bibitem{patel2019understanding}
M.~Patel, J.~S. Kim, H.~Hassan, and O.~Mutlu, ``{Understanding and Modeling
  On-Die Error Correction in Modern DRAM: An Experimental Study Using Real
  Devices},'' in \emph{DSN}, 2019.

\bibitem{patel2020beer}
M.~Patel, J.~Kim, T.~Shahroodi, H.~Hassan, and O.~Mutlu, ``{Bit-Exact ECC
  Recovery (BEER): Determining DRAM On-Die ECC Functions by Exploiting DRAM
  Data Retention Characteristics},'' in \emph{{MICRO}}, 2020.

\end{thebibliography}

\end{document}